\documentclass[final,1p,times,twocolumn,floatfix]{elsarticle}

\usepackage{amssymb}
\usepackage{amsthm}
\usepackage{amsmath}
\usepackage{amsbsy}
\usepackage{nicefrac}

\journal{Astroparticle Physics}

\begin{document}

\maketitle

\begin{frontmatter}

%
%

\title{Antarctic Radio Frequency Albedo and Implications for Cosmic Ray Reconstruction}
%
%

%
%

\author[label0,label13]{D. Z. Besson} 
\author[label0]{J. Stockham} 
\author[label0]{M. Sullivan} 
\author[label1,label4]{P. Allison}
\author[label3]{S. W. Barwick}
\author[label4]{B. M. Baughman}
\author[label4]{J. J. Beatty}
\author[label2]{K. Belov} 
\author[label6]{S. Bevan} 
\author[label7]{W. R. Binns} 
\author[label8]{C. Chen} 
\author[label8]{P. Chen} 
\author[label9]{J. M. Clem} 
\author[label4]{A. Connolly}
\author[label9]{D. De Marco} 
\author[label7]{P. F. Dowkontt} 
\author[label1]{M. DuVernois} 
\author[label3]{D. Goldstein} 
\author[label1]{P. W. Gorham} 
\author[label4]{E. W. Grashorn} 
\author[label1]{B. Hill} 
\author[label2,label11,label11a]{S. Hoover}
\author[label10]{M. Huang}
\author[label7]{M. H. Israel}
\author[label9]{A. Javaid}
\author[label1]{J. Kowalski}
\author[label1]{J. Learned}
\author[label10]{K. M. Liewer}
\author[label1]{S. Matsuno}
\author[label4]{B. C. Mercurio}
\author[label1]{C. Miki} 
\author[label6]{M. Mottram}
\author[label3,label8]{J. Nam}
\author[label10]{C. J. Naudet}
\author[label6]{R. J. Nichol}
\author[label4]{K. Palladino}
\author[label1,label10]{A. Romero-Wolf}
\author[label1]{L. Ruckman} 
\author[label2]{D. Saltzberg}
\author[label9]{D. Seckel}
\author[label8]{R. Y. Shang}
\author[label0]{M. Stockham} 
\author[label1]{G. S. Varner}
\author[label2,label12]{A. G. Vieregg}
\author[label8]{and Y. Wang}

\address[label0]{Department of Physics and Astronomy, University of Kansas, Lawrence, KS 66045, USA}
\address[label1]{Department of Physics and Astronomy, University of Hawaii at Manoa, Honolulu, HI 96822, USA}
\address[label2]{Department of Physics and Astronomy, University of California, Los Angeles, CA 90095, USA}
\address[label3]{Department of Physics, University of California, Irvine, California 92697, USA}
\address[label4]{Department of Physics, Ohio State University, Columbus, Ohio 43210, USA}
\address[label6]{Department of Physics and Astronomy, University College London, London, United Kingdom}
\address[label7]{Department of Physics, Washington University in St. Louis, Missouri 63130, USA}
\address[label8]{Department of Physics, National Taiwan University, Taipei, Taiwan 10617}
\address[label9]{Department of Physics, University of Delaware, Newark, Delaware 19716, USA}
\address[label10]{Jet Propulsion Laboratory, Pasadena, California 91109, USA}
\address[label11]{Kavli Institute for Cosmological Physics, University of Chicago, 5640 South Ellis Avenue, Chicago, IL 60637, USA}
\address[label11a]{Enrico Fermi Institute, University of Chicago, 5640 South Ellis Avenue, Chicago, IL 60637, USA}
\address[label12]{Harvard-Smithsonian Center for Astrophysics, Cambridge, Massachusetts 02138, USA}
\address[label13]{Moscow Engineering and Physics Institute, 31 Kashirskaya Shosse, Rossia 115409}

\begin{abstract}
We describe herein a measurement of the Antarctic surface ``roughness'' performed by the 
balloon-borne ANITA (ANtarctic Impulsive Transient Apparatus) experiment. Originally purposed for cosmic-ray astrophysics, the
radio-frequency (RF) receiver ANITA gondola, from its 38 km altitude vantage point, can scan a disk of snow surface
600 km in radius. The primary purpose
of ANITA is to detect RF emissions from cosmic rays incident on Antarctica, 
such as neutrinos which penetrate through the atmosphere and interact within the ice, resulting in signal directed upwards which then refracts at the ice-air interface and up and out to ANITA, or high-energy nuclei (most likely irons or protons), which interact in the upper atmosphere (at altitudes below ANITA) and produce a spray of down-coming RF which reflects off the snow surface and back up to the gondola. The energy of such high-energy nuclei can be inferred from the observed reflected signal only if the surface reflectivity is known. We describe herein an attempt to quantify the Antarctic surface reflectivity, using the Sun as a constant, unpolarized RF source. 
We find that the
reflectivity of the surface generally follows the expectations from the Fresnel equations, lending support to the use
of those equations to give an overall correction factor to calculate cosmic ray energies for all locations in Antarctica.
The analysis described below is based on ANITA-II data.
After launching from McMurdo Station in December, 2008, ANITA-II was
aloft for a period of 31 days with a typical instantaneous
duty cycle exceeding 95\%. 
\end{abstract}

%
%

%


%
%
\end{frontmatter}

\section{Introduction}
Antarctic surface features are expected to be macroscopic, at typical length scales of order 0.01-1 meter, and can therefore be explored using radiation with a comparable wavelength. We can probe surface roughness with radio wave receivers sensitive over a similar wavelength regime, by measuring the ratio of the intensity of the surface-reflected radio-frequency Solar image to the Solar image observed directly by the balloon-borne ANITA experiment. That measurement can then be compared to the ratio expected for reflection off of a smooth surface (``specular'' reflection). 
By taking the ratio of the surface-reflected Solar RF power to the direct Solar RF power measured with ANITA, as a function of incident elevation angle relative to the surface, $\theta_i$, we can thus estimate the surface power reflection coefficients ${\cal R}(\theta_i)$. 

\section{\label{IntroANITA}The ANITA Experiment}
Initiated in 2003, the Antarctic Impulsive Transient Antenna (ANITA) is a balloon-borne antenna array primarily designed to detect radio wave pulses caused by neutrino collisions with deep Antarctic ice \citep{gorham_2009a}. The basic instrument consists of a suite of 40 quad-ridged horn antennas, optimized over the frequency range 200-1200 MHz, with separate outputs for vertically vs. horizontally-polarized incident radio frequency signals, mounted to a high-altitude balloon. From an elevation of $\sim$38 km, the payload observes the Antarctic continent in a circumpolar trajectory. 
Two one-month long missions (ANITA-I; Dec. 2006-Jan. 2007 and ANITA-II; Dec. 2008-Jan. 2009) have yielded world's-best limits on the flux of Ultra-High Energy (``UHE'', corresponding to energies in excess of $10^{20}$ eV) neutrinos\citep{gorham_2007,gorham_2010}. Nevertheless, model-dependent calculations of surface roughness effects show considerable variation in the predicted fraction of signal power emerging upwards from within the dense ice target medium and across the ice-air interface to the receiver, and therefore incur significant uncertainties in our neutrino energy estimates; this is also true for experiments seeking to measure neutrino interactions within the lunar regolith\citep{Lunaska,GLUE,RESUN,Puschino} producing RF emerging from within the lunar surface and subsequently detected at Earth. 

Although it yielded no definitive neutrino detections,
an interferometric analysis of the ANITA-I data sample (unexpectedly) provided a statistically large (16 events) sample of radio frequency signals attributed to the geomagnetic + Askaryan radiation associated with cosmic-ray induced extensive air showers\citep{hoover_2010} (EAS). Those events were supplemented with an additional four EAS detections during the ANITA-II flight two years later.
The ANITA-III launch, scheduled for December, 2014, is expected to result in an improvement by a factor of $\sim$100 in the number of detected air shower events relative to ANITA-I. Correspondingly, we have sought to determine the surface reflectivity in advance of that upcoming ANITA-III mission.

\subsection{ANITA Data Acquisition and Signal Reception Chain}
\begin{figure}[htpb]
\centerline{\includegraphics[width=0.6\textwidth]{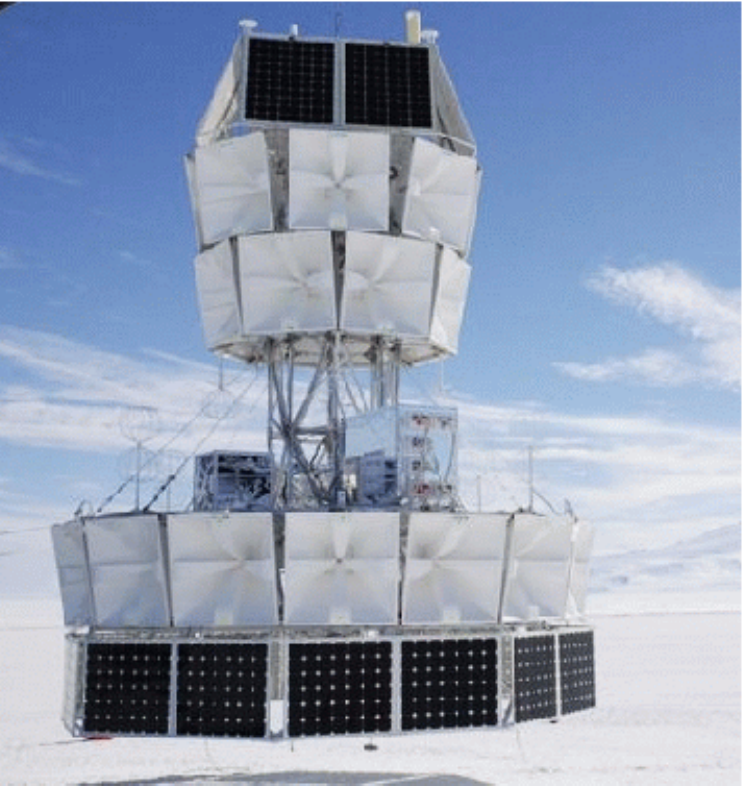}}
\caption{The ANITA payload, prior to launch from McMurdo Station, Antarctica. Three `rings' of dual-polarization Seavey `horn' antennas are evident in Figure; above and below the horn antennas are rows of solar panels capable of providing nearly 1 kW to the ANITA array. Note that the antennas are slightly `canted' at 10 degrees below the horizontal to maximize the cosmic ray sensitivity.}
\label{fig:ANITA}
\end{figure}

The ANITA detector is designed to efficiently collect radio signal 
over a 1 GHz bandwidth, with electronics and trigger configured to produce the minimum possible energy threshold.
The front-end Seavey quad-ridge horn antennas (Figure \ref{fig:ANITA}) provide the first
element in the ANITA radio wave signal processing chain. These horn antennas
have separate VPol
vs. HPol signal polarization readouts, and each has a field-of-view
of approximately $25^\circ$ half-width-half-maximum (HWHM) in 
both horizontal azimuth angle ($\phi$) and vertical elevation angle ($\theta$). 
Following the Seavey antennas, high- and low-pass filtering restricts the system bandpass
to 200--1200 MHz, after which
signals are split into a `trigger' path
consisting of a hierarchy of trigger conditions
(``levels'') and a `digitization' path. 
If all levels
of the three-tiered, hierarchical trigger are satisfied, the digitized signals, consisting of 256 voltages sampled for each antenna at 2.6 GigaSamples/s, are stored for later analysis.
During typical ANITA-II data-taking conditions, the rate at which all three tiers of the trigger are satisfied is approximately 10 Hz; i.e., 
roughly every 100 ms, 100 ns of voltage vs. time data, for both polarizations of each antenna, are stored to disk.
The overwhelming majority of these triggers are due to upward excursions of the thermal radio photosphere.

\subsection{ANITA Radio Air Shower Observations}
Radio frequency emissions from extensive air showers were observed both during the ANITA-I (16 total events) and ANITA-II (4 events) flights. An analysis of the characteristics of the observed air showers indicated that the cosmic ray progenitors were within an order of magnitude of the highest energy particles ever observed at Earth.
Interestingly, two of the ANITA-II events featured reflection points consistent with sea ice, for which the dielectric contrast relative to the air, and therefore the expected reflected power, is expected to be larger compared to reflections from sheet ice.

\subsection{Relationship between albedo and cosmic ray reconstruction}
The cosmic rays to which ANITA is sensitive are among the highest energy particles observed to date, 
with energies $10^7$ times higher than those capable of being generated in our terrestrial
accelerators. The primary physics interest is in the nature of the cosmic accelerators
producing such objects and specifically how they achieve the
extraordinary energies measured by ANITA. The surface roughness, which, for the purposes of this paper, is inferred by the radio frequency  
albedo, therefore has direct consequences for ANITA's estimate of the energy of their observed cosmic rays. Numerically, given an observed radio-frequency signal power $P_{signal}$, the 
surface-reflectivity corrected cosmic ray energy $E_{CR}$ is just given by an overall scale factor and is proportional to
\begin{equation}
 E_{CR}\propto P_{signal}/(F(\theta_i)f_{Roughness}), 
\end{equation}
where $F(\theta_i)$ is the expected surface reflectivity for a smooth surface,
$f_{Roughness}$ is the fraction of signal strength reflected relative to the expectation for ``perfect''
scattering from a smooth surface (i.e., $f_{Roughness}\le$1). In our experiment, we are sensitive to the product
\begin{equation}
 {\cal R}=(F(\theta_i)f_{Roughness}).
\end{equation}
\subsection{Geometry of Measurements and Definition of Terms}
The roughness can be quantified using the Rayleigh criterion.
According to this definition, the magnitude of variation in surface height $h$,
as measured along the line of the incoming RF plane wave must satisfy the condition
$h\cos(\phi_L)<0.125$ for 
the surface to be considered `smooth'; correspondingly `smoothness' is 
largest for radar signals looking directly `down' into surface ridges, rather than viewing `across' those same ridges
($h$ here represents the ground's height variation, 
while the look angle, $\phi_L$, is measured from the nadir to the RF direction of travel). In the 
terminology used above, $f_{Roughness}$ decreases with increasing $\theta_i$.

\begin{figure}[htpb]
\centerline{\includegraphics[width=0.6\textwidth]{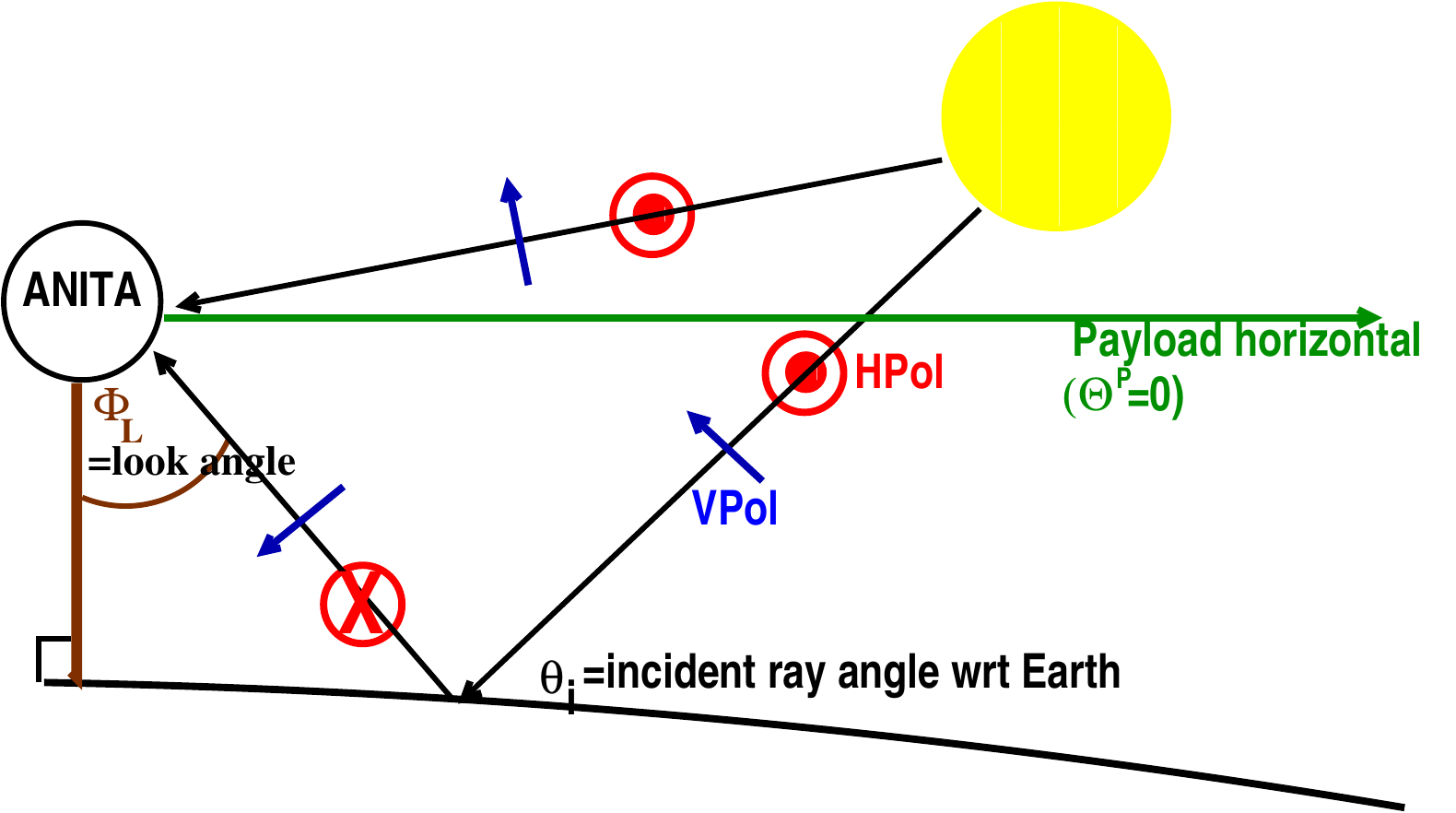}}
\caption{Geometry of measurements presented herein, illustrating Solar reflection elevation angle $\theta_i$, as well as conventions for VPol and HPol signal polarizations.}
\label{fig:geom}
\end{figure}

Figure \ref{fig:geom} illustrates some of the angle conventions used for the remainder of this document. We separately refer to $\theta^P$, as measured from the payload horizontal, and for which the Earth's horizon corresponds to $\theta^P=-6^\circ$, and $\theta_i$, which is most relevant for our reflection measurements, and is referenced relative to a tangent at the surface location of the Solar reflection point. Also shown in the Figure is the look angle $\phi_L$, which is, as discussed previously, related to the surface roughness.

\begin{figure}[htpb]
\centerline{\includegraphics[width=0.6\textwidth]{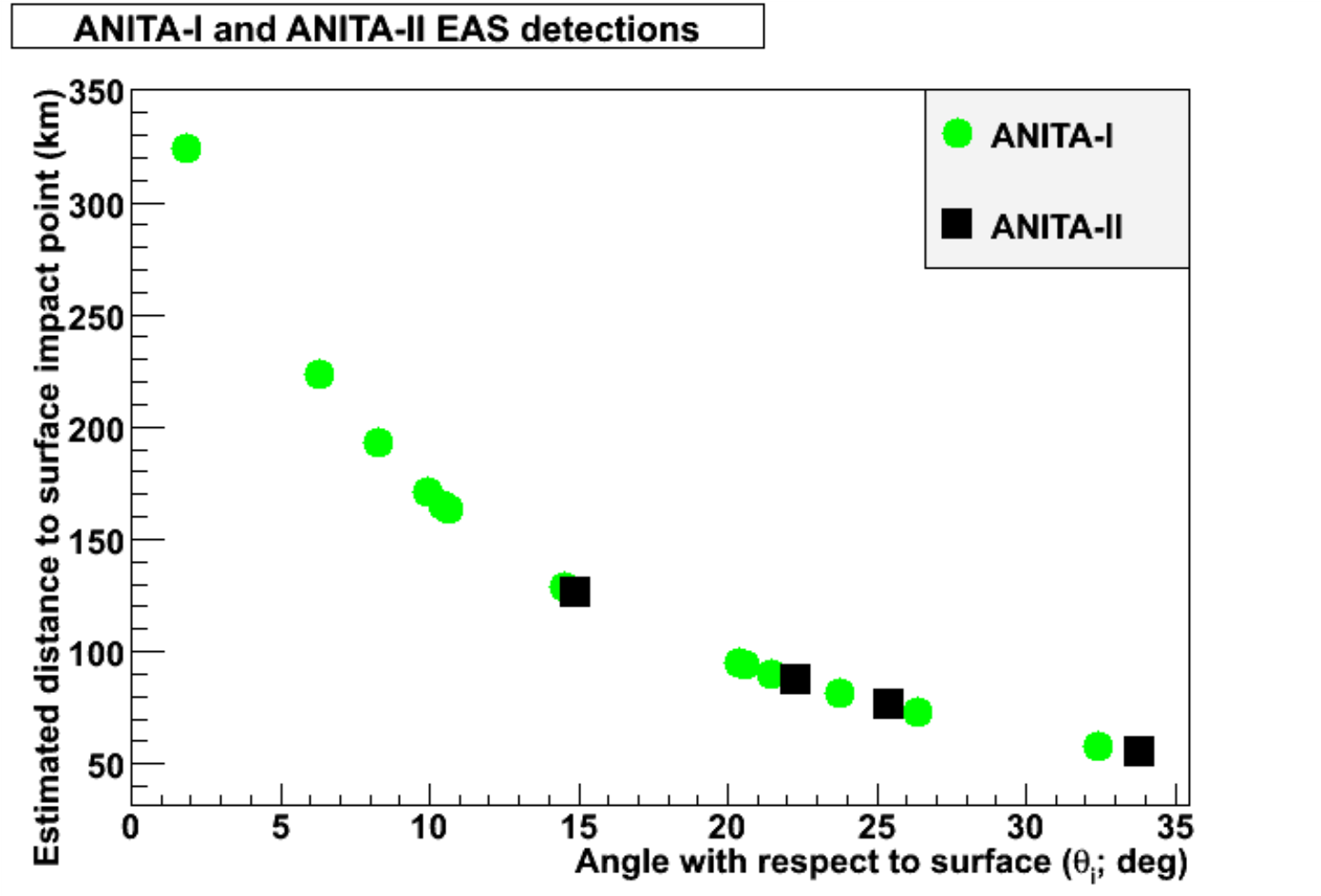}}
\caption{Elevation angle $\theta_i$, relative to the horizon, for the 14 ANITA-I, and 4 ANITA-II cosmic ray events observed via their surface-reflected radio emissions. Indicated distances are estimated knowing the GPS coordinates (latitude, longitude and elevation) of the gondola for each of the ANITA events shown, and extrapolating back to the source `location' on the surface.}
\label{fig:RFangles}
\end{figure}

Using the coordinate convention followed in this paper for comparing intensity measurements to expectations from the Fresnel equations, the incidence angles $\theta_i$ for the ANITA radio-reflected EAS sample\citep{hoover_2010} are shown in Figure \ref{fig:RFangles}. 
From simple solid angle considerations ($d\Omega\sim\cos\theta_i d\theta_i$), we expect the acceptance to decrease with increasing incident elevation angle. Additionally, since the antennas are canted at $-10$ degrees relative to the payload horizontal, we have poor sensitivity to sources at large values of $\theta_i$. Correspondingly, the best experimental sensitivity to EAS is 
in the angular range $\theta_i<$25 degrees.

\section{\label{SurfaceRefl}Antarctic Surface Reflectivity}

We now consider what is known
and what measurements have been made thus far. We note that there are no previous measurements
of the Antarctic reflectivity over the ANITA bandpass, and, in fact, no previous angle-dependent
measurements as comprehensive as ours over any frequency.

\subsection{Visible Frequencies}
The polar surface reflectivity (alternately, ``albedo'') is one of the crucial inputs to climate models, given its importance in determining the thermal
equilibrium between Earth and Sun. Since the Solar power spectrum peaks in the visible,
measurements of Solar reflectivity have typically focused on the 3900-6600 $\AA$ regime\citep{WarrenBrandt98}. 
At such scales, sastrugi effects dominate, as a function of look angle $\phi_L$ relative to the sastrugi 
alignment. Long-term studies (particularly at South Pole Station) have established periodic variations in
the Solar albedo in the visible wavelength interval, over $\sim$5 year time scales\citep{WangZender11}.
Such measurements can also be used as a proxy for wind direction measurements\citep{ParishBromwich87}, to the extent that albedo observations probe wind-blown sastrugi orientation.

\subsection{Measurements at GHz Frequencies}
\begin{figure}[htpb]
\centerline{\includegraphics[width=0.6\textwidth]{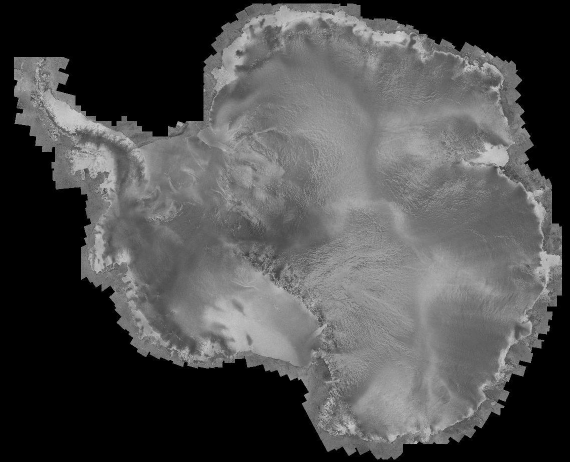}}
\caption{Measured RADARSAT Ku-band reflected signal $\sigma_0$. Maximum contrast corresponds to a variation of almost 20 dB in measured reflected power (based on publicly available data in http://bprc.osu.edu/rsl/radarsat/data/). No corrections have been made for the relative orientation of transmitter receiver or the orientation of the satellite during data-taking.}
\label{fig:RadarSat}
\end{figure}

The Earth has, to date, been extensively mapped in the Ku-band (12-18 GHz) \citep{KeithStuart12} and also in the lower-frequency C-band (4--8 GHz), most recently, and most comprehensively by the Canadian-based RADARSAT satellites (http://bprc.osu.edu/rsl/radarsat/data/). These relatively high frequencies, and correspondingly small wavelengths compared to the typical scale of surface features, probe not only pure `surface' features, but also sub-surface volumetric scattering due to, e.g., buried sastrugi.
The absolute(/relative) RADARSAT signal return strength calibration of 2 dB(/1 dB) is sufficient to map the reflected power across the Antarctic continent (Figure  \ref{fig:RadarSat}), which exhibits variations of order 20 dB.
Those results are qualitatively consistent with the Envisat satellite
Ku-band reflection measurements\citep{Frederique}, albeit with some large deviations, particularly around the Ross Ice Shelf, where the reflectivity may be dominated by effects not present in the larger Antarctic ice sheet. Currently, the available Envisat
data sample is being extended to the S-band (3.2 GHz central frequency) as well as the Ka-band (37 GHz central frequency)\citep{Fred1}.

\begin{figure}[htpb]
\centerline{\includegraphics[width=0.6\textwidth]{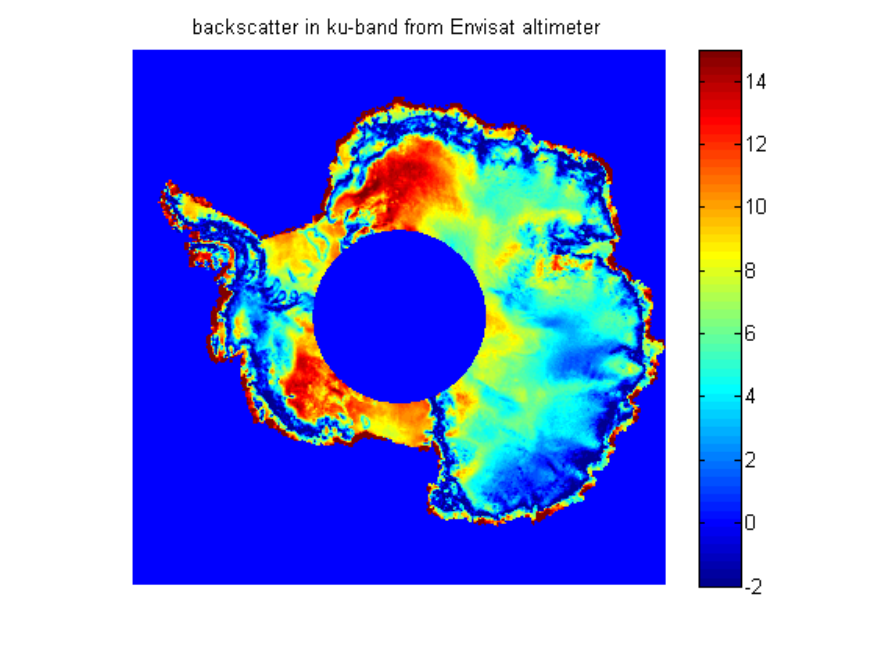}}
\caption{Envisat Ku-band measured reflected signal\citep{Frederique}, in units of dB. No corrections have been made for the relative orientation of transmitter/receiver or the orientation of the satellite during data taking. Figure courtesy of Frederique R\'emy.}
\label{fig:Frederique}
\end{figure}

The Envisat C-band reflection data
generally follow the Ku-band trend shown in Figure \ref{fig:Frederique}, but with the measured
C-band reflected power over 
West Antarctica reduced by $\sim$2-5 dB; the data in East Antarctica is in
general agreement with the Ku-band data to within 2-3 dB.

Although the RADARSAT nor Envisat surveys do not correlate their observations with expectations from the Fresnel coefficients, assuming that the satellite observations are made at a constant incidence angle and polarization, their results may be interpreted as probing regions of large surface roughness (corresponding to weak returns resulting from diffuse scattering) and smoother surface scattering (corresponding to stronger, more coherent returns). This interpretation is consistent with the fact that both satellite surveys observe considerably smaller reflected signal strength near the Antarctic coast, where generally higher winds should result in larger surface features, as well as the Trans-Antarctic Range, which is presumably characterized by irregular, exposed rock. Indeed, an overlay of these data with an Antarctic wind velocity map shows striking similarities. 
Comparable data are not readily available at the 30 cm--1.5 m wavelengths of interest to the ANITA mission.



\section{Interferometry Applied to Solar Imaging}
\subsection{ANITA interferometry}
Beginning with the eighty 100-ns duration voltage
waveforms which are taken during an ANITA event trigger (40 antennas $\times$ 2 polarizations per antenna),
familiar radio interferometry techniques are used to identify localized radio sources. To determine the direction of an external impulsive source, the full solid angle is first divided into an $N_\theta\times N_\phi$ grid (typically, $180\times 360$) in elevation and azimuth, respectively, corresponding to pixels of solid angle one square degree. Next, an inter-channel cross-correlation sum is calculated for each point in the grid, by shifting cross-correlated (upsampled by a factor of two to improve precision) waveforms by the relative propagation time delays that would be expected, channel-by-channel, for a putative source at that ($\theta$,$\phi$) grid point. Formally, the cross-correlation ${c_{ij}}$, for two channels $i$ and $j$ can be written as the dot product of the voltage time series for channel $i$: $V_i(t)$, and the voltage time series for channel $j$: $V_j(t)$, shifted by the calculated time delay $\delta t_{ij}^S(\theta,\phi)$ corresponding to the putative source $S$ at ($\theta$,$\phi$), for the $N$ samples in the dot product. For example, if $\delta t_{ij}$=5 ns, then, at the upsampled value of 5.2 GigaSamples/s, the relative shift in waveforms is equal to 31 bins, leaving a total of $N$=512-31=481 overlapping samples remaining to calculate the dot product. To minimize sensitivity to channel-dependent gain variations, this cross-correlation coefficient is normalized relative to the rms voltages in the two channels $\sigma(V_i)$ and $\sigma(V_j)$:
\begin{equation}
c_{ij}=\sum_{n=1}^NV_i^S(t_n)V_j^S(t_n+\delta t_{ij}^S)/\sqrt{\sigma(V_i)\sigma(V_j)}
\end{equation}
In the limit of a strong source which dominates the 100-ns waveform, the cross-correlation rises linearly with signal voltage $V^S$, since $\sigma(V)\sim V^S$. In the limit of a very weak source, $\sigma(V)$ is largely independent of $V^S$ and the cross-correlation increases quadratically with signal voltage. The high signal-to-noise ratio (SNR) events that comprise the ANITA EAS sample correspond to the former case. The Solar signal described below corresponds to the latter case, and therefore, in the case of signals from the Sun, the cross-correlation coefficient scales with the radio power received. 
The total cross-correlation ${\cal C}$ is calculated by summing over all independent $(i,j)$ pairs, for antennas having overlapping beam patterns. For each grid pixel, the voltages from nine antennas (one central antenna and eight neighbors) are used to form the
total cross-correlation. For each event,
the polar and azimuthal angles of a putative source are then assigned to that particular ($\theta,\phi$) grid pixel, out of the $180\times 360$ tested, which gives the highest total cross-correlation ${\cal C}$.
More extensive details on the construction of the 
ANITA interferometric map can be found in the literature\citep{AndresPaper}.

\begin{figure}[htpb]
\centerline{\includegraphics[width=0.6\textwidth]{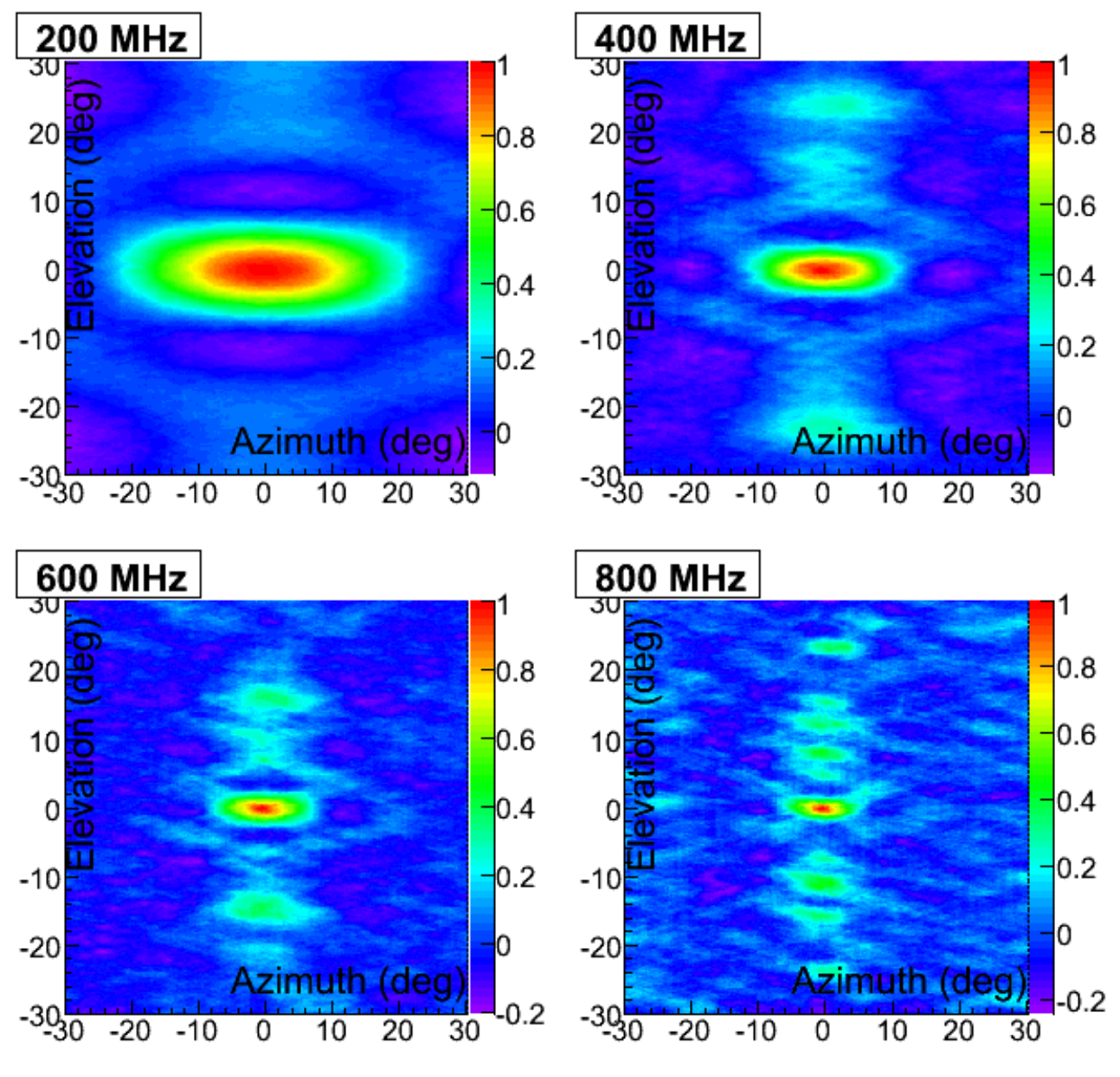}}
\caption{Interferometric map for simulated 200, 400, 600, and 800 MHz continuous-wave source at ($\theta^P_S,\phi_S$=0,0). Each pixel in this 360$\times$180 ($\phi\times\theta$) pixel map represents the cross-correlation strength ${\cal C}$ and assuming a source at infinity at that value of $\phi$ and $\theta$ to determine the appropriate pair-wise channel-by-channel time delays. The normalization of the signal here is arbitrary; the signal-to-noise ratio has been set to an arbitrarily high value so as to illustrate the qualitative features of the signal.}
\label{fig:cw.pdf}
\end{figure}

Interferometry is particularly suited for 
reconstruction of 
continuous-wave (CW) sources radiating at a fixed value of frequency.
Monte Carlo simulations, in which a known source (either CW or transient) at a known location illuminates our simulated receiver array, can be used to investigate interferometric reconstruction techniques. For our simulated source, we can calculate the expected time delay for a radio-frequency signal arriving from that known location, and generate synthetic signal ``waveforms'' which are then superimposed on a model of the background, presumed to be thermal noise. These synthetic signal data can then be reconstructed using the same algorithms applied to real ANITA data.
As illustrated by these Monte Carlo simulations (Figure  \ref{fig:cw.pdf}), the interferometric reconstruction resolution improves, roughly linearly, with the frequency of the source. Alternately, the angular size of the interferometric image provides information on the source frequency content, with wider images implying power spectra weighted preferentially to lower frequencies. 

A solar-powered discone transmitter, buried in a 15-cm diameter borehole approximately 100 m into the ice sheet at a remote site near Taylor Dome, Antarctica in December 2006,
has been used to assess the inteferometric source reconstruction precision for in-ice signals with broad frequency content. For that transmitter, we obtain resolutions of 0.25 degrees in elevation and 0.56 degrees in azimuth for the ANITA-II flight. By comparison, the intrinsic Solar radio source size is only slightly larger than the optical Solar image, i.e., of order 1 square
degree in the sky.

\subsection{Solar RF emissions}
\label{sec:Solar}
In the absence of bursts or flares, the Sun conveniently provides the ANITA mission an uninterrupted, unpolarized, (relatively)
constant power calibration source. However, since the Solar emissions in the ANITA bandpass are typically sub-threshold relative to the ANITA trigger requirements, the Solar image is only clearly visible after adding many interferometric images in an heliocentric coordinate system. 
For this analysis, we co-add the interferometric maps for $\sim 10^3-10^4$ events, excluding triggers containing high signal-to-noise ratio sources, to elucidate the Solar signal. In order that the Solar signal not be smeared by the motion of the ballon or sun over the corresponding data collection time, the interferometric maps are summed only after making the appropriate event-by-event corrections, pixel-by-pixel, in both azimuth and also polar angle.

To illustrate the expected signals,
Figures \ref{fig:e10} and \ref{fig:e35} 
show simulated interferometric maps, for which a single point source, with a flat frequency spectrum 
has been modeled at the indicated elevation angle. 
Although reflection effects were not modeled in constructing these maps, we nevertheless observe some fringing, characteristic of ANITA
interferograms, and resulting from the regular lateral or vertical spacing
of the ANITA horn receiver antennas, evident in Figure \ref{fig:ANITA}.
We note that for a source at a payload elevation $\theta^P_S$, fringing tends to produce an enhancement at an elevation
angle typically a few degrees below $-\theta^P_S$ and therefore contributes some contamination to our measured Solar
reflection power, with a magnitude that grows with $\theta^P_S$, requiring a modest correction to our raw, extracted reflection signals.

Figure  \ref{fig:Sun} shows the image of the Sun both in vertical and horizontal polarizations in data, 
after co-adding 10000 events, acquired over a time of approximately 20 minutes. During that time, the Sun typically moves 0.5 degrees in the sky in elevation and 12 degrees in azimuth.
As the Sun has no intrinsic polarization, we expect approximately equal source intensities in each polarization, consistent with observation. 

\subsection{Cross-check of measurement procedure}
We have conducted three cross-checks to verify the veracity of the interferometric technique, as applied to the Sun, as follows.

\subsubsection{Variation of Solar signal strength with time}For a stationary summer observer at latitudes within the Antarctic circle, the Sun is expected to trace a sinusoid across the sky over a diurnal period. Since the Seavey horn antennas used by ANITA are canted at 10 degrees below the horizon, the sun is apparently brightest when it is lowest in the sky. We observe an obvious 24-hour cycle of observed Solar source intensity (Figure \ref{fig:SolarSignalVTime}), as tracked for an 8-day period of the flight following the balloon's launch from McMurdo Station.

\subsubsection{Check of Observed Solar Power Spectrum}
The raw Solar power spectrum over a wide spectral range is shown in Figure \ref{fig:SolarSpectrum} (taken from the British Astronomical Organization at http://www.britastro.org/radio/RadioSources/sun.html); the curve of greatest relevance for this measurement corresponds to the ``quiet sun'', in the 25 cm$\to$1.25 m wavelength interval.
Figure \ref{fig:SolarObservatories} shows the absolute spectral power measured at two terrestrial observatories, Learmonth (http://www.ips.gov.au/World{\_}Data{\_}Centre/1/10) and SOLRA (http://radiosun.ts.astro.it/eng/solra.php), 
averaged over the time period comprised by the ANITA-II mission. Overlaid on that graph is the direct Solar power, as measured interferometrically by ANITA-II. 
We observe reasonable agreement in the shape of the two spectra.
\message{Z: What is the timescale over which the 'ambient' Solar power changes?}
\message{jess: working on compiling this for the two observatories}

\subsubsection{Antenna Gain vs. Elevation}Over the course of a typical one-month Antarctic Balloon mission, the Sun's elevation in the sky, relative to the payload ``horizontal'' $\theta^P$, varies between 10 and 40 degrees, and therefore allows an in-flight determination of the polar beam pattern for the ANITA dual-polarization quad-ridged horn antennas, which have been calibrated in the laboratory and found to have approximately $\pm$25-degree (sigma) beam aperture. Figure \ref{fig:BeamGainPattern} overlays the dependence of the peak interferometric Solar power (defined as the magnitude of the highest-amplitude pixel in the Solar interferometric map) on Solar elevation, compared with the expectation using a Gaussian function, with standard deviation as indicated, to represent the horn response as a function of off-axis angle. We observe that the Solar interferometric map can be used, in-flight, to monitor the off-axis antenna response to an accuracy of a few degrees.

\section{Measurement of Surface Radio Reflectivity ${\cal R}$}
We now describe the primary result of this paper, namely the measurement of the Solar signal observed via its Antarctic surface reflection, compared to the Solar signal measured directly. In principle, this is predicted for smooth, or ``specular'' reflections by the Fresnel coefficients. We have therefore attempted to quantify the degree to which our reflected signal deviates from the expectation from specular reflection, which itself indicates effects due to surface non-homogeneity.
\subsection{Flight Path and Surface Reflection Locations}
The ANITA flight path samples the interior ice sheet, as well as shelf ice and some sea ice at latitudes North of the Ross Ice Shelf. To illustrate the region of Antarctica sampled by the ANITA trajectory, Figure  \ref{fig:latlongRefl} shows the latitude and longitude of the gondola (indicated by red circles) and also the direction along the ice surface of the Solar reflection location (small black lines), calculated using the known gondola and Solar positions at the time of each event trigger, and taking into account Earth curvature effects.  

\subsection{Quantitative Technique}
Similar to Figure \ref{fig:Sun}, we can create an enhanced radio frequency reflected Solar image by co-adding many events, including corrections for i) balloon motion, ii) Earth motion, and iii) Earth surface curvature effects.
Although the location of the Solar reflection on the surface is obviously closely correlated to the direct Solar motion in azimuth, and also closely anti-correlated to the direct Solar motion in elevation, corrections due to local Earth curvature, surface elevation and the viewing angle relative to the surface can easily smear the reflection image by an amount exceeding the intrinsic expected Solar beam spot size ($\sim$1 degree) and must therefore be carefully tracked as the reflection images are stacked. We note that large-scale variations in surface slope gradients in Antarctica typically occur over hundreds of km, and constitute, at most, a 0.1 degree smearing for the analysis presented below.

\message{from abby:  Figure 13: get rid of titles.  This may be true for some other plots, too. - Figure 13: What is the lighter colored curve on the bottom? }
For this analysis, to determine the reflected surface power, we fit the direct and reflected Solar signal peaks, defined by projecting the data in Figure \ref{fig:Sun} satisfying $|\phi|<$5 degrees onto the elevation ($\theta$) axis. Signals are fit
to Gaussian functions, with the integral under our Gaussian signal parametrization taken as our primary estimate of the signal power. The use of a Gaussian is motivated by the fact that it approximates the off-axis beam response of the ANITA horn antennas. The functional form chosen for the solar power signal fit is $A(\theta)=A_0{\rm exp}(-(\theta-\theta_0)/2\sigma^2_\theta)$, with $A_0$ the amplitude of the fitted Gaussian, $\theta_0$ the central value of the fitted peak (approximately zero for the direct solar signal), 
and $\sigma_\theta$ the antenna beam width in this particular bin of $\theta$. 
Figures \ref{fig:HPolFit} and \ref{fig:VPolFit} show fits to the one-dimensional projections of the reflection-stacked interferometric maps corresponding to Figure  \ref{fig:Sun}, from which we extract signal estimates. For simplicity, we have centered the Solar image in the center of the graphs shown.

As a cross-check, we have also performed a second, independent analysis independent of Gaussian fitting, which uses the maximum amplitude pixel in the interferometric map, in the vicinity of the known Solar, or reflected location in the sky, as a measure of the signal intensity. 
This technique works well for cases of high SNR, which is true for the HPol reflected signal, but becomes decreasingly reliabile for low SNR since it is more subject to bin-to-bin fluctuations and binning effects, in comparison to signal extractions which fit the signal over several bins. Correspondingly, we have only applied this cross-check to the HPol case, for which the signal is strongest.

Figure  \ref{fig:FF.pdf} displays the average ratio ${\cal R}$ of the Solar radio reflection power relative to the direct Solar image, for both vertical as well as horizontal antenna polarizations, including all antenna gain corrections, as a function of the reflection angle of the Solar image off the Earth's surface, relative to the balloon. 
Also shown in Figure  \ref{fig:FF.pdf} are the expected Fresnel reflection power coefficients, assuming specular reflection off the Antarctic surface, and taking the index of refraction of the surface snow to be $n_{surface}=1.35$ from the one direct measurement of this value\citep{Krav2004} at radio frequencies; this value is also consistent with estimates derived from Antarctic surface snow density\citep{Schytt56}.

We note that the vertical spread of points relative to the Fresnel expectation in Figure \ref{fig:FF.pdf} is approximately a factor of three larger than the vertical spread of the `direct-Sun' points shown in Figure \ref{fig:BeamGainPattern}, indicating that the systematic errors in extracting the signal amplitude of the weaker surface reflection are considerably larger than the signal estimation uncertainties inherent in the interferometric maps themselves.

Figure \ref{fig:FF1.pdf} displays the average value, and the rms spread in values, for each bin of incidence angle presented in Figure \ref{fig:FF.pdf}, and more clearly illustrates the deviation of the VPol signal from zero at all angles. The tendency of the reflection coefficients to `fall' as $\theta_i$ approaches oblique incidence angles is also evident from this Figure. To illustrate the statistical quality of our data, we have investigated the extent to which our distribution of residuals (i.e., measured reflectivity minus the Fresnel reflection coefficient expected for a given incident angle) is consistent with a normal distribution. Figure \ref{fig:ResidualFit.pdf} shows the result of fitting the HPol deviations to a Gaussian bell curve and shows that the distribution of residuals is approximated satisfactorily by such a distribution.

\subsection{Cross-check: Comparison of reflected intensity over sea ice vs. shelf ice}
The reflected signal strength at an interface depends on the dielectric contrast between the air and the reflecting layer. The Solar image reflected off sea ice, having relatively higher salt content, should be brighter than the Solar image reflected off sheet ice. As illustrated in Figures \ref{fig:HIceWater} and \ref{fig:VIceWater}, we correspondingly note the largest enhancement in the reflected signal strength in the coastal region beyond the Ross Ice Shelf, consistent with expectation.

\subsection{Interpretation}
Overall, we observe reasonable agreement between the experimental points and the Fresnel expectation for specular reflection. However, we observe an apparent excess, most pronounced for VPol, of the received signal relative to the expectation from the Fresnel equations, at higher $\theta_i$ angles. In particular, we do not observe the predicted signal dimunition (approaching zero signal strength) as $\theta_i$ approaches the Brewster angle of 36 degrees. We attribute this to some combination of a) the result of surface roughness effects, which result in non-zero signal at the Brewster angle, b) possible under-accounting of the effects of interferometric fringing, c) possible effects due to the asymmetry in the ANITA-II event trigger, which has higher sensitivity to thermal excursions of VPol over HPol, and d) an under-accounting of the cross-polarization isolation (i.e., the degree to which pure VPol signals induce a response in the HPol ports of the Seavey antenna, and vice versa). Such cross-polarization effects are, in principle, accounted for in Figure \ref{fig:FF.pdf}, but with the assumption that the cross-polarization is constant as a function of frequency and off-boresight angle. Our results suggest that isolation may be poorer at large off-boresight angles and increasing with frequency.

\section{Discussion and Summary}
To determine the energy of a typical down-coming air shower event measured by its radio emissions, the ANITA experiment requires knowledge of the RF reflectivity of the Antarctic ice surface. The reflectivity is, in turn, closely related to the physical scale of surface features, which depend on such parameters as local wind speed and direction and surface elevation gradient. Surface roughness also has implications for neutrino detection, as high-frequency, small-wavelength radio waves resulting from in-ice neutrino interactions may decohere as they pass through the ice-air interface, en route to detection at the ANITA gondola. The wavelength regime to which ANITA is sensitive (30 cm$\to$150 cm) approaches the expected size of surface inhomogeneities, and therefore underscores the importance of direct surface roughness measurements. 

In our current analysis, the ANITA-II balloon-borne radio interferometer has been used to measure the Antarctic surface reflectivity over the ANITA passband. 
For $\theta_i>15^\circ$, we find general consistency with the values of ${\cal R}(\theta_i)$ expected from the Fresnel equations\citep{Hecht02}, which prescribe the amount of signal power reflected at the interface between two smooth dielectrics given their indices of refraction, for both vertical- vs. horizontal-polarizations (``VPol'' and ``HPol'', respectively). At more glancing incident angles ($\theta_i < 15^ \circ$), our data suggest slightly reduced signal strength compared to the expectation from the Fresnel equations, perhaps indicating that surface roughness effects are becoming increasingly apparent at oblique incidence angles.
Quantitatively, we find, overall, reasonable agreement (typically, of order 10-15\%), for both Vertical as well as Horizontally polarized signal power, in the average measured reflectivity, when compared to expectations based on direct application of the Fresnel equations. 

Although a direct comparison of our results with the Ku-band and C-band surveys is complicated by the much more limited surface coverage of this study, as well as the fact that our results are solar-geometry specific, the general consistency of our reflection coefficients with the Fresnel expectation, to within $\pm$2 dB is markedly smaller than the $\sim$20 dB variations reported by those higher-frequency measurements, indicating that the typical length (wavelength) scales at which surface decoherence sets in as a result of inhomogeneities is of order, but not more than, 10 cm. An analysis of S-band (2-4 GHz) reflectivity data may therefore be the most promising frequency band in which to search for surface roughness effects.

As an indication of what the associated systematic errors in our cosmic ray energy estimates might be, Figure \ref{fig:dErg} presents the relative reflectivity uncertainty, for a given ultra-high energy cosmic ray event, implied by our analysis. The value shown at each viewing angle value is equal to the standard deviation of the values, about the Fresnel expectation, and are therefore directly transcribed from Figure \ref{fig:FF1.pdf}. Thus, for example, we expect the uncertainty in the overall energy estimate of a cosmic ray observed at incidence angles up to 30 degrees to have (at maximum, given the low SNR of the Solar event sample we have used for our analysis) a contribution due to surface reflectivity effects which does not exceed 35-40\%. This is the first numerical estimate of the ANITA energy uncertainty.

We stress, however, that for the much higher signal-to-noise typical of EAS events, signal energy estimates can be made directly from the event waveforms, without relying directly on the interferometric maps to infer signal strength\citep{hoover_thesis}. Variations in the event-by-event energy estimate should be considerably smaller than those encountered in the Solar reflection analysis; the systematic errors implied by Figure \ref{fig:dErg} therefore represent a (very) conservative upper limit. 
This analysis therefore provides some basis for directly applying the Fresnel coefficients to infer the energies of cosmic rays measured by their surface-reflected radio-frequency emissions, as well as the energies of neutrinos detected via their emissions across the air-ice interface. Ongoing analysis will incorporate these results, combined with improved simulations of the radio frequency EAS signal, to give revised estimates of the observed ANITA-I cosmic ray events. 


Studies of the frequency-dependence of the Solar reflection, as well as correlations with known wind patterns across the continent, are currently in progress. In parallel, improved interferometric techniques based on triple-, rather than pair-wise waveform correlations are being developed, with the goal of suppressing confusion introduced by fringe effects (as described in this paper) and enhancing the interferometric signal-to-noise ratio.


%
%
%
%
%
%
%

\centerline{Acknowledgments}

We express our gratitude to Dr. Frederique R\'emy for kindly 
providing us with the Envisat reflectivity data.
We also thank the National Aeronautics and Space Administration,
the National Science Foundation Office of Polar Programs,
the Department of Energy Office of Science HEP Division,
the UK Science and Technology Facilities Council, the
National Science Council in Taiwan ROC, Fermilab's QuarkNet Program, 
the Russian Ministry of Science and Education,
and especially the
staff of the Columbia Scientific Balloon Facility.
A. Romero-Wolf would like to thank NASA (NESSF Grant NNX07AO05H) 
for support for this work. 
This material is based upon work supported by the National Science Foundation Graduate Research Fellowship under Grant No. NSF0064451. Data used for this analysis and to generate the included figures may be obtained by contacting the corresponding author, J. Stockham $(jegab8@ku.edu)$.

%
\begin{figure}[htpb]
\centerline{\includegraphics[width=0.6\textwidth]{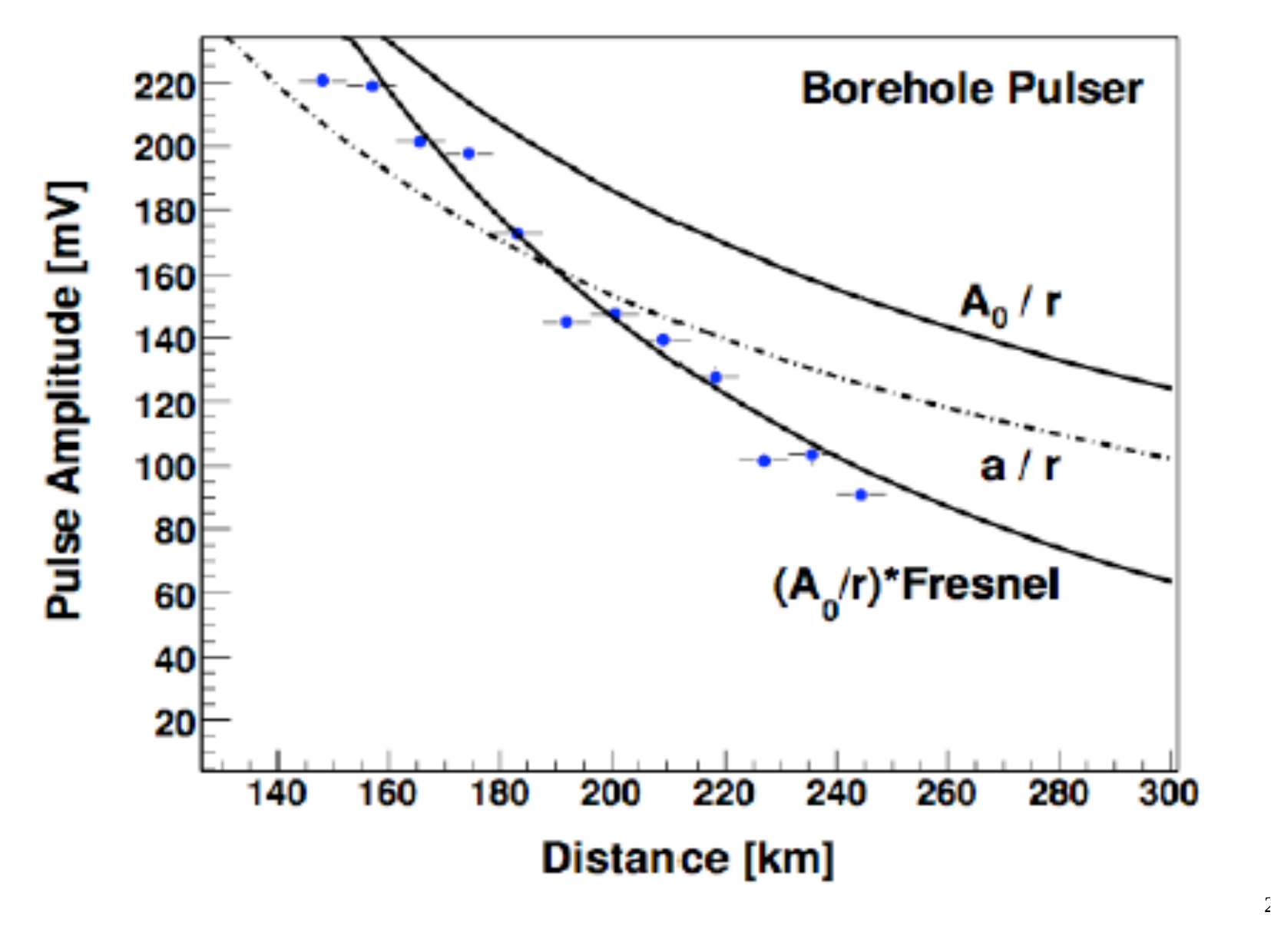}}
\caption{Measured signal voltage received from ANITA-I englacial calibration pulser (data points), compared to signal voltages calculated for three different cases: a) amplitude scaling assumed as $r^{-1}$ and assuming a spherical transmitter beam, without inclusion of Fresnel factors (labeled ``a/r''), b) same as a), but taking into account the actual beam pattern of the englacial discone transmitter (labeled ``$A_0$/r''), and c) same as b) with inclusion of estimated Fresnel transmission coefficient across a smooth ice-air interface (``($A_0$/r)*Fresnel''). The agreement between the blue points and the smooth-reflection Fresnel-model curve indicate that surface roughness effects are not substantial at that site.}\label{fig:TDGoldstein}\end{figure}
\begin{figure}[htpb]
\centerline{\includegraphics[width=0.6\textwidth]{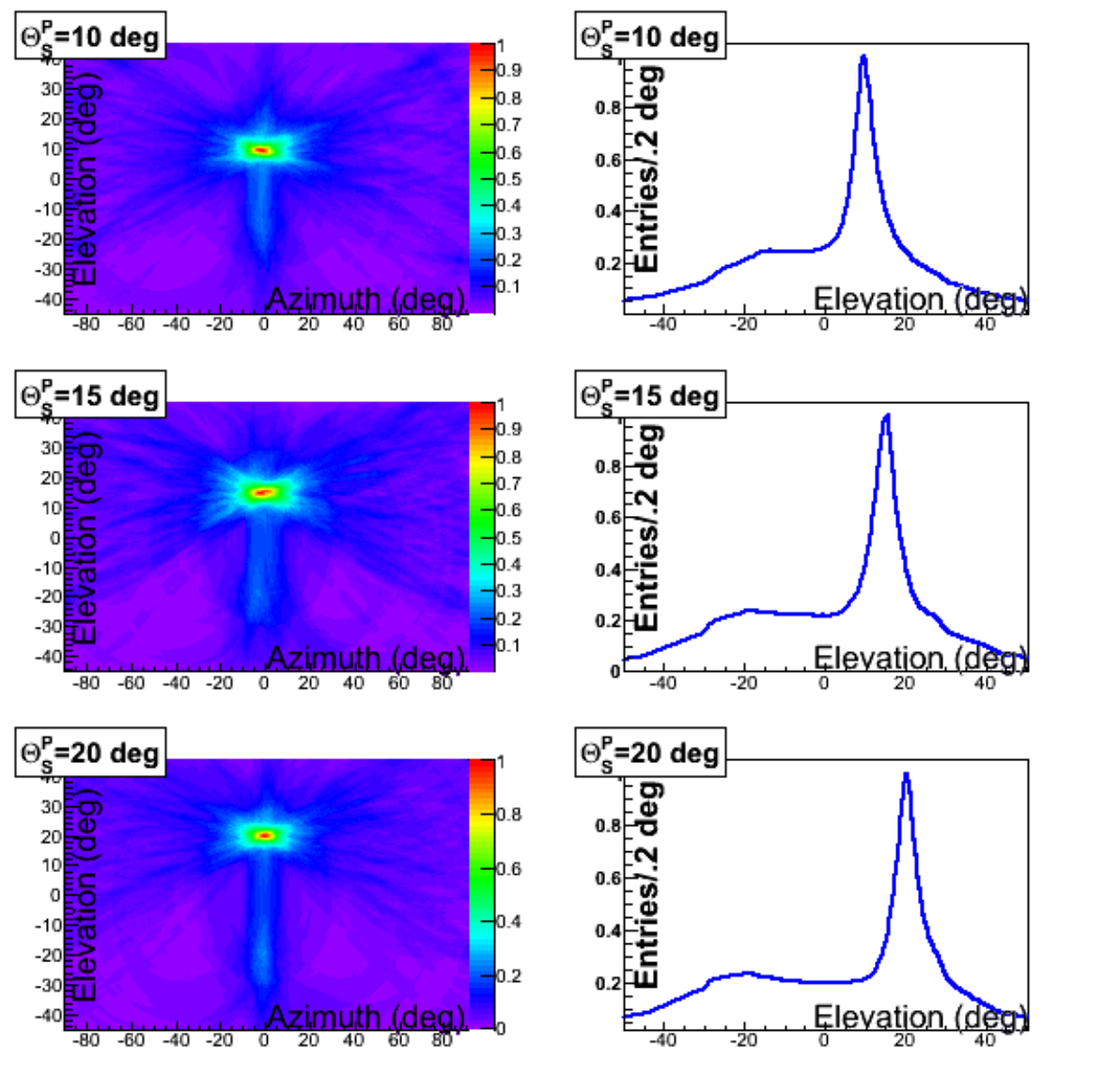}}
\caption{Interferometric map for simulated source at 10, 15, and 20 degree payload elevations, and projections onto elevation axis (right). Color scale has units of $Voltage$. For these simulations, the simulated source is considerably smaller than one pixel. Although no surface reflection effects are present in the simulation, note the presence of a spurious interferometric `fringe' at a nearly equal elevation angle below the horizon.}
\label{fig:e10}
\end{figure}
\begin{figure}[htpb]
\centerline{\includegraphics[width=0.6\textwidth]{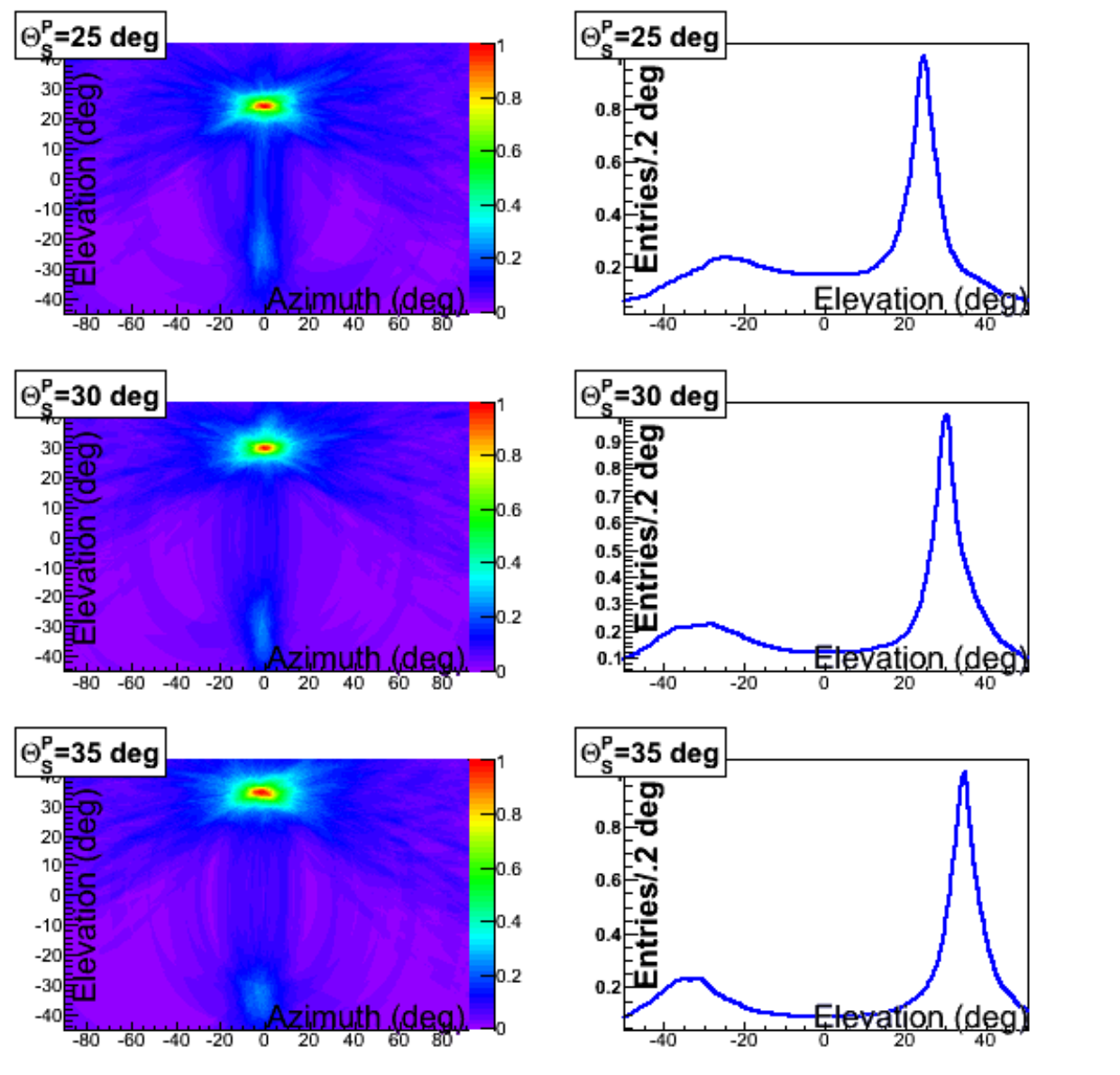}}
\caption{Interferometric map for simulated source at 25, 30, and 35 degree payload elevations (left), and projections onto elevation axis (right). Color scale has units of $Voltage$.}
\label{fig:e35}
\end{figure}
\begin{figure}[htpb]
\centerline{\includegraphics[width=0.6\textwidth]{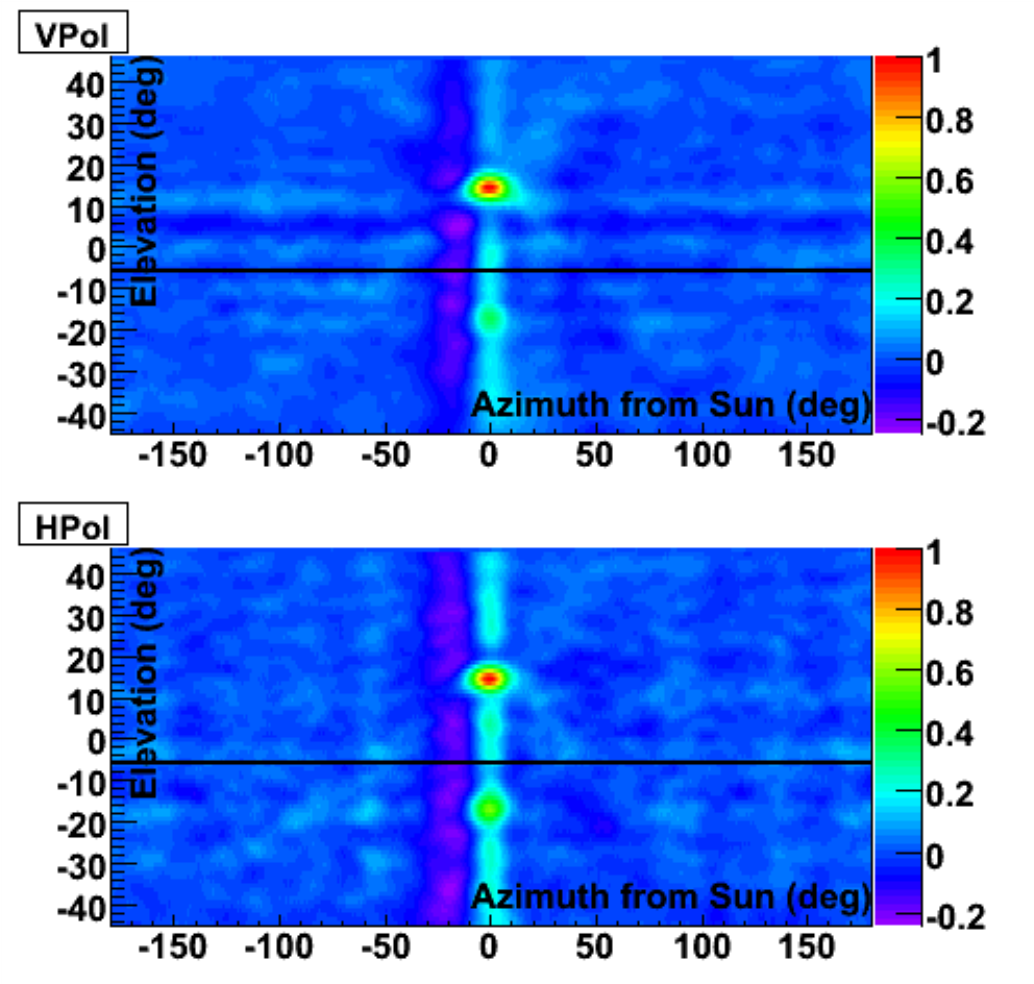}}
\caption{Interferograms (top, VPol and bottom, HPol) formed by tracking the location of the Sun for 10000 events. Normalization is identical for both top and bottom plots. Both direct and surface-reflected Solar signals are evident, at angles of $\pm 14^\circ$. For this plot, we use payload coordinates, for which the horizontal plane is defined as $\theta^P=0^\circ$, with elevation angles towards the zenith/(nadir) defined as positive/(negative). Interferograms are stacked to account for both the motion of the Sun, as well as the translational motion of the ANITA gondola across the Earth's surface and the rotational motions of ANITA relative to its vertical and horizontal axes. The ice/air horizon, relative to the balloon, is indicated by the solid black line at $\theta^P=-$6 degrees. Angles lower than this correspond to in-ice sources; angles larger correspond to in-air sources.} 
\label{fig:Sun}
\end{figure}
\begin{figure}[htpb]
\centerline{\includegraphics[width=0.6\textwidth]{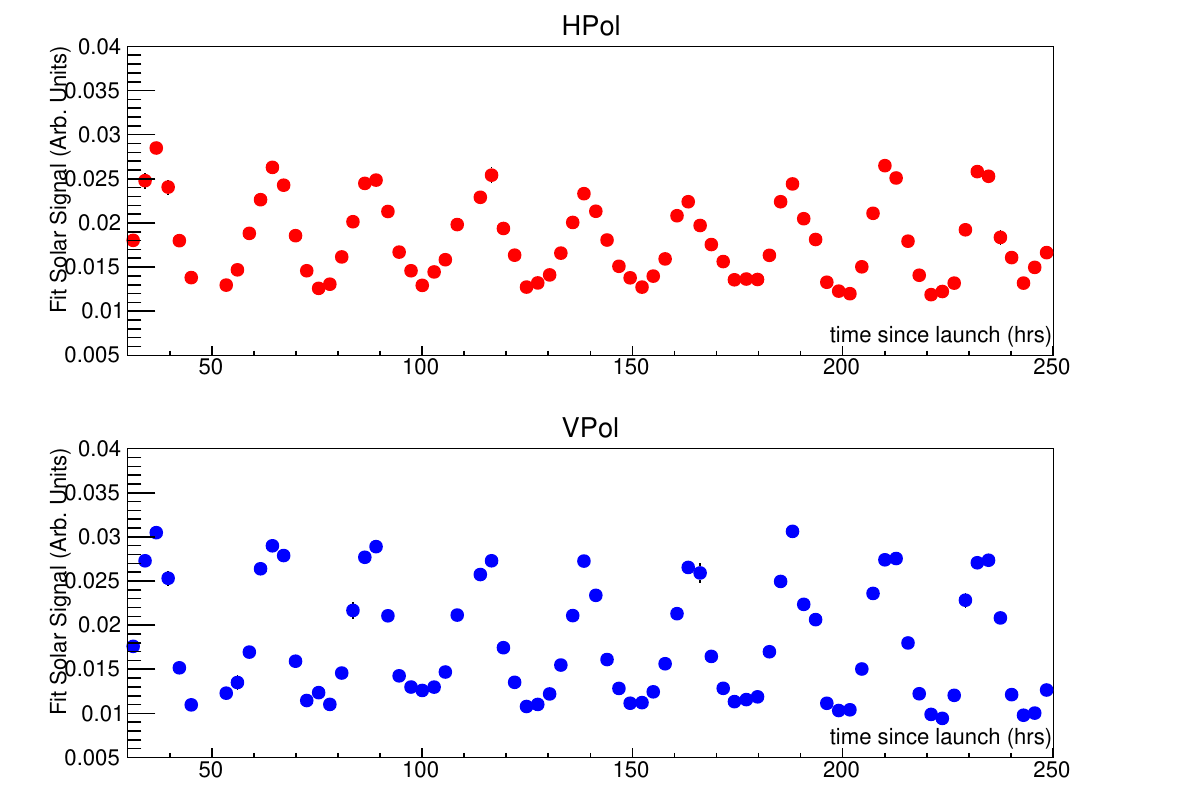}}
\caption{Solar signal strength as a function of number of hours into flight. Normalization is identical for both top and bottom plots. As expected, we observe a diurnal variation, as the solar location moves in elevation relative to the ANITA receiver antenna array beam center.}
\label{fig:SolarSignalVTime}
\end{figure}

\begin{figure}[htpb]
\centerline{\includegraphics[width=0.6\textwidth]{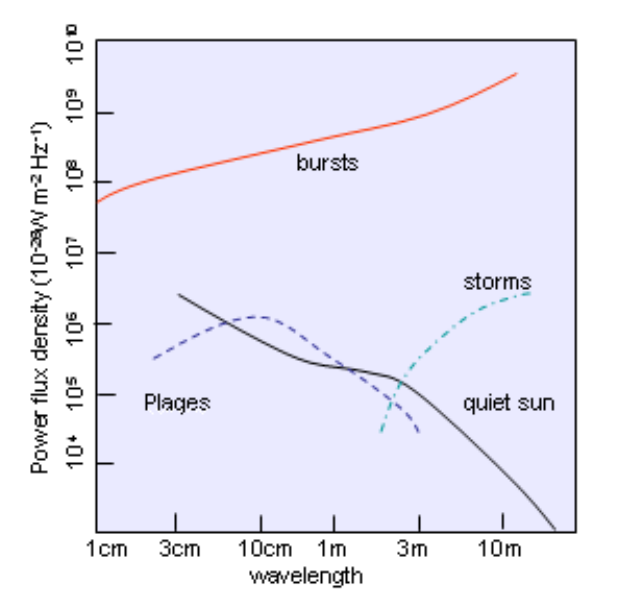}}
\caption{Raw solar power spectrum, as a function of emission wavelength, compiled from non-ANITA data. During the Dec, 2008--Jan, 2009 period of the ANITA-II flight, the sun was in its `quiet' phase.}
\label{fig:SolarSpectrum}
\end{figure}
\clearpage
\begin{figure}[htpb]
\centerline{\includegraphics[width=0.6\textwidth]{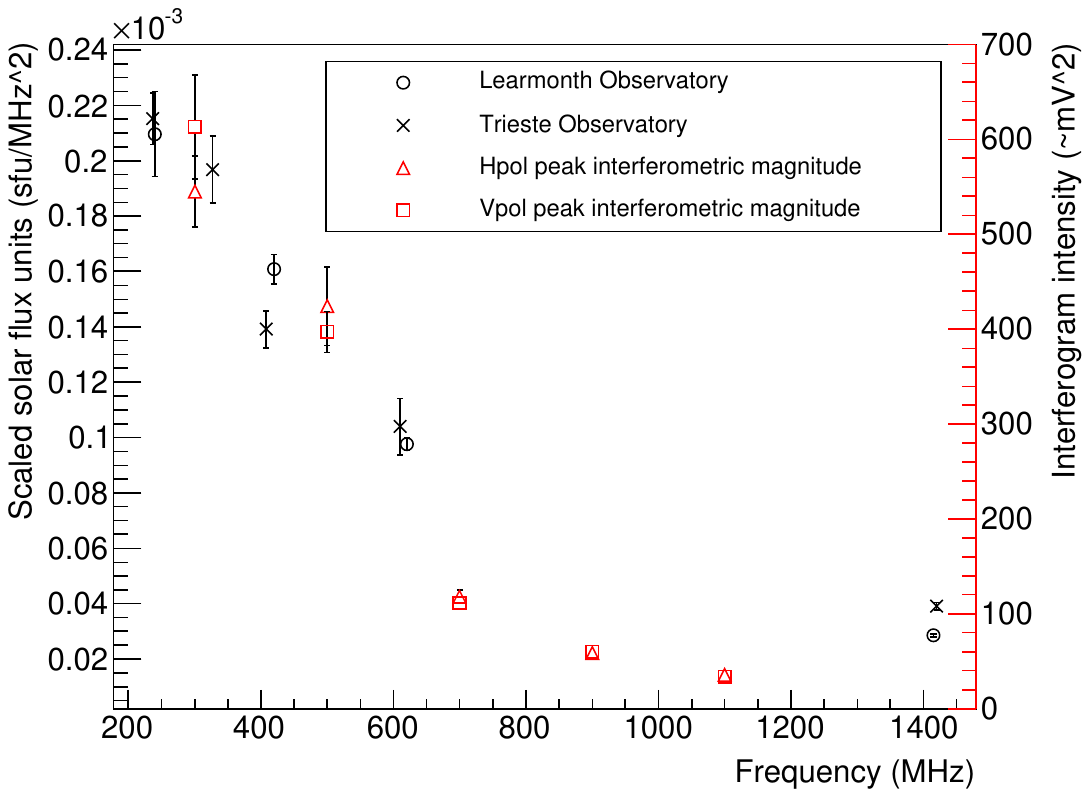}}
\caption{Solar intensities measured at two Solar observatories in Learmonth, Australia (open circles) and Trieste, Italy (crosses) compared to the frequency-banded Solar intensities retrieved from the interferograms (Hpol, open triangles and Vpol, open squares). The Trieste data are corrected for latitude of the observations. The ANITA data are obtained from interferograms filtered into five bands (200-400 MHz, 400-600 MHz, 600-800 MHz, 800-1000 MHz, 1000-1200 MHz); points are placed at the center frequency of each band. For each band, we use the same sample of 8000 co-added event triggers. The ANITA data have been scaled to match the terrestrial observatory data at 300 MHz. Overall, the trend we observe from the ANITA data tracks the observations from the ground-based Solar observatories.} 
\label{fig:SolarObservatories}
\end{figure}
\begin{figure}[htpb]
\centerline{\includegraphics[width=0.6\textwidth]{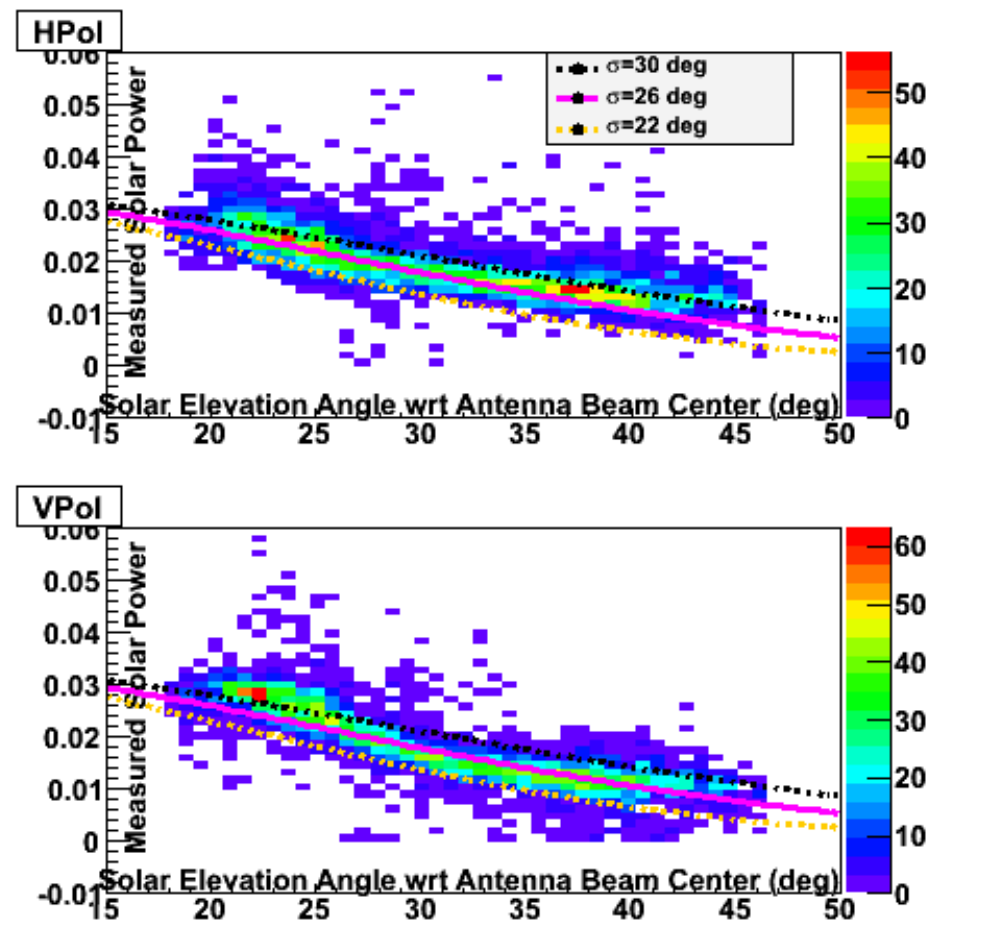}}
\caption{Parametrized antenna gain of varying beam width (lines) compared to relative Solar intensities extracted from Hpol (top) and Vpol (bottom) interferograms. Color scale indicates number of events observed at a given point in graph. Legend indicates expectations for measured signal intensity, as a function of Gaussian antenna beam width $\sigma$. We find best agreement with data using a beam width of $\sigma\sim 26^\circ$; this value is then used to correct the observed magnitude of either the direct Solar signal or the surface-reflected Solar signal, assuming a Gaussian beam width profile. The nominal ANITA antenna beam width for the Seavey horns, as calibrated pre-flight, varies between 25--30 degrees, depending on frequency and E-plane/H-plane polarization.}
\label{fig:BeamGainPattern}
\end{figure}
\begin{figure}[htpb]
\centerline{\includegraphics[width=0.6\textwidth]{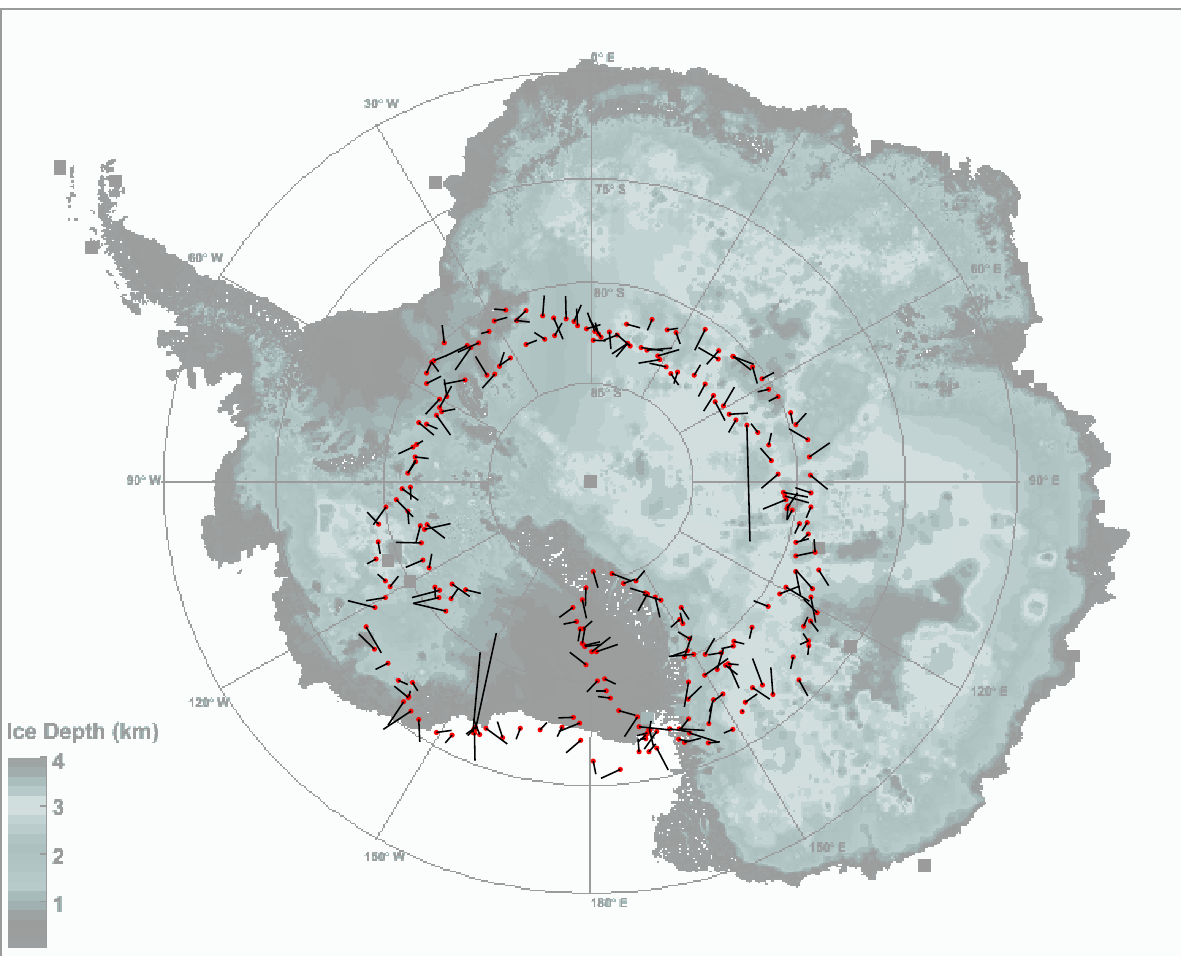}}
\caption{Illustration of typical geometry of ANITA-II gondola relative to Solar surface reflection point. ANITA-II locations sampled during flight are shown for single events. Red points correspond to balloon location; each associated black line represents the vector pointing from the ANITA location to the Solar reflection location on the ice surface. Ice thickness values are taken from BEDMAP.}
\label{fig:latlongRefl}
\end{figure}
\begin{figure}[htpb]
\centerline{\includegraphics[width=0.6\textwidth]{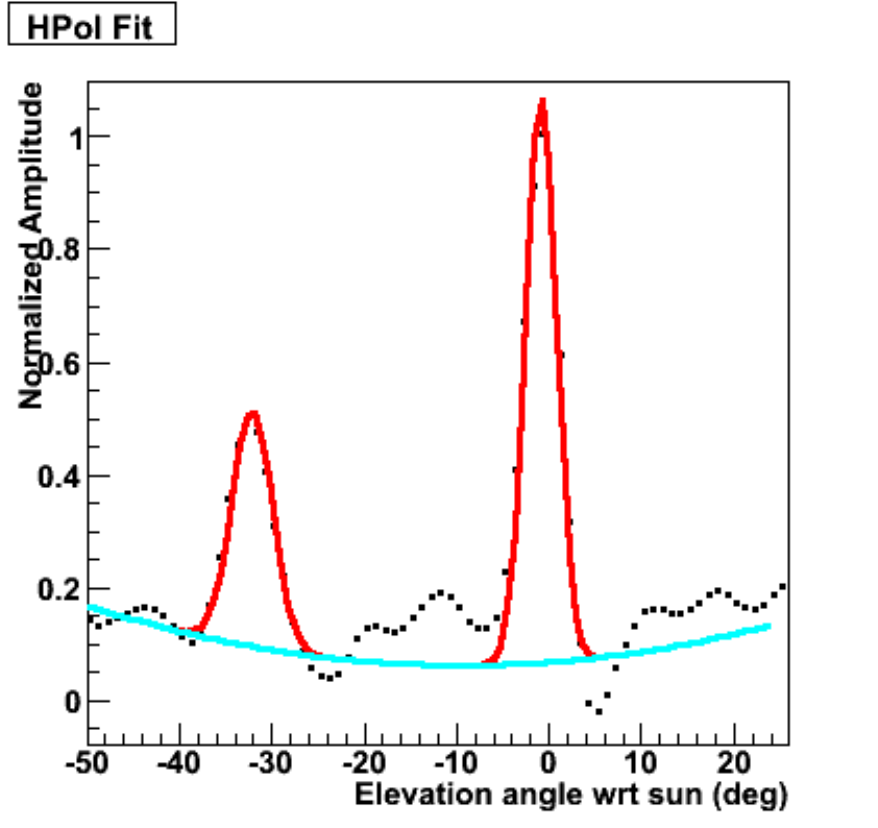}}
\caption{Sample fit to projection of HPol, Sun-centered Solar interferometric map onto elevation axis, after restricting our interferometric map to $\pm$10 degrees around $\phi=0$ so as to isolate the signal region. Fit includes two Gaussian-shaped signals, corresponding to direct Solar signal and reflected surface reflection, plus a second-order polynomial to approximate non-zero background under Gaussian signals. The relative signal intensity is taken as the integral of the appropriate Gaussian signal (either direct Solar or surface-reflected Solar).}
\label{fig:HPolFit}
\end{figure}
\begin{figure}[htpb]
\centerline{\includegraphics[width=0.6\textwidth]{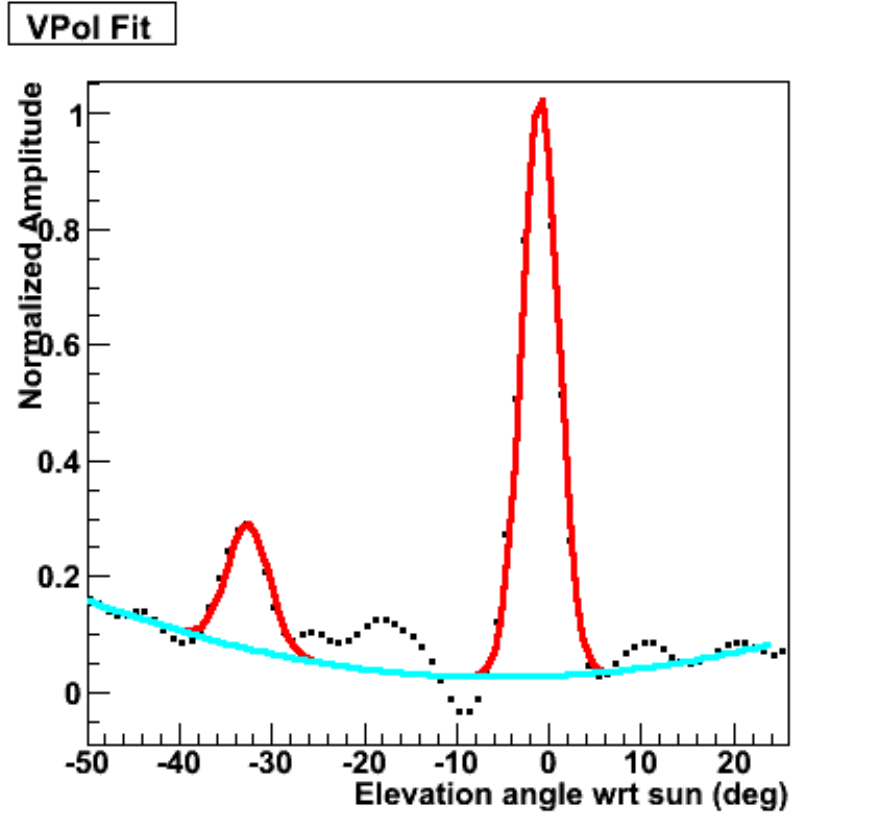}}
\caption{Sample fit of projection of VPol, Sun-centered interferometric map onto elevation axis, after once again cutting $\pm$10 degrees around $\phi=0$. Fit includes two Gaussian-shaped signals, corresponding to direct Solar signal and reflected surface reflection, plus second-order polynomial to approximate non-zero background under Gaussian signals.}
\label{fig:VPolFit}
\end{figure}
\begin{figure}[htpb]
\centerline{\includegraphics[width=0.6\textwidth]{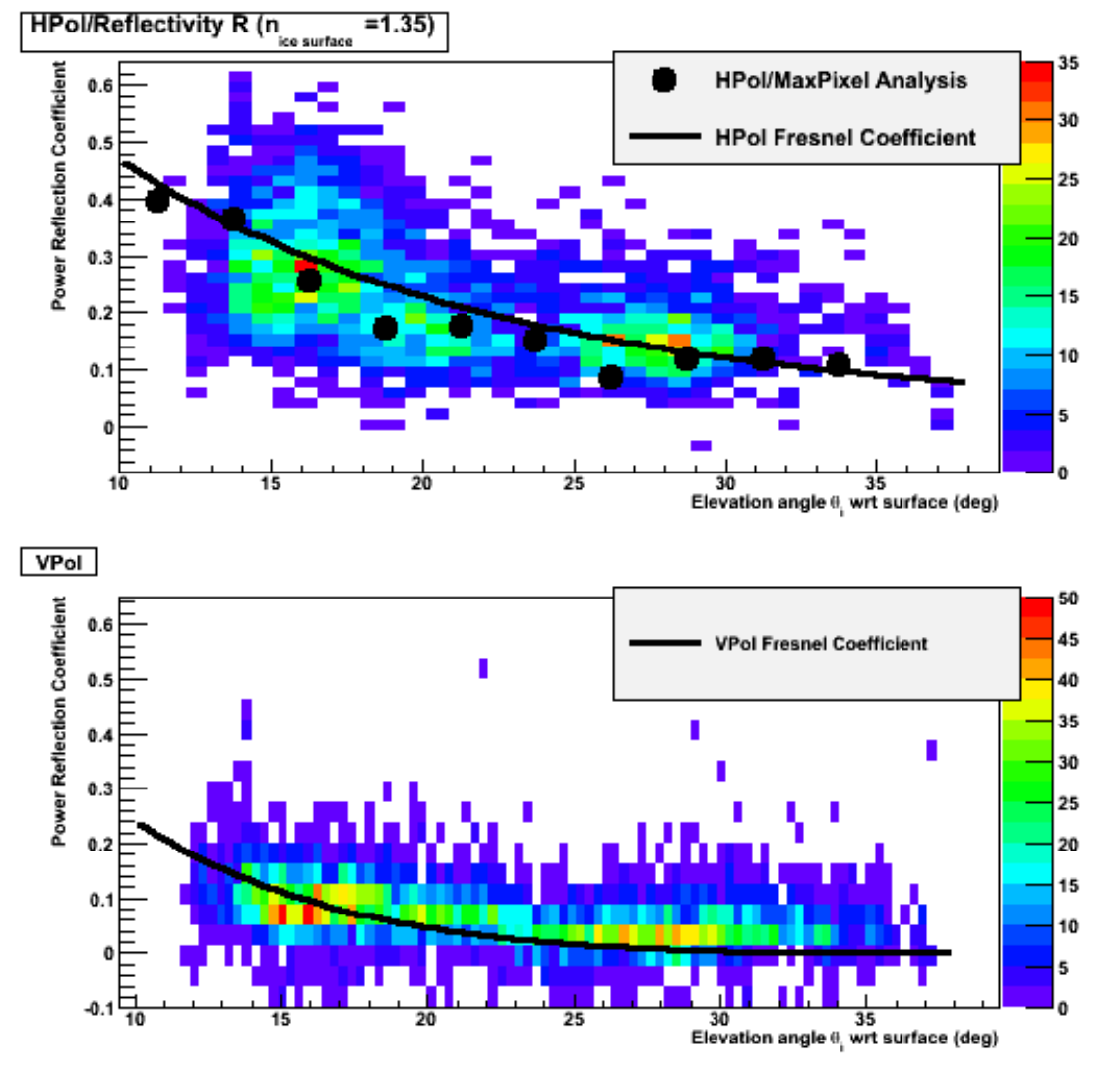}}
\caption{Ratio of ANITA-II measured reflected Solar power to directly-received Solar power, as a function of incident elevation angle, for both HPol and VPol antennas. Overlaid are also calculated values as obtained from direct application of the Fresnel equations to calculate the expected reflected signal power ${\cal R}$. Color scheme indicates number of 10-minute data samples yielding indicated measured reflection coefficient, for a given value of $\theta_i$; in total, there are approximately 5000 data points in each of the upper and lower plots.}
\label{fig:FF.pdf}
\end{figure}
\begin{figure}[htpb]
\centerline{\includegraphics[width=0.6\textwidth]{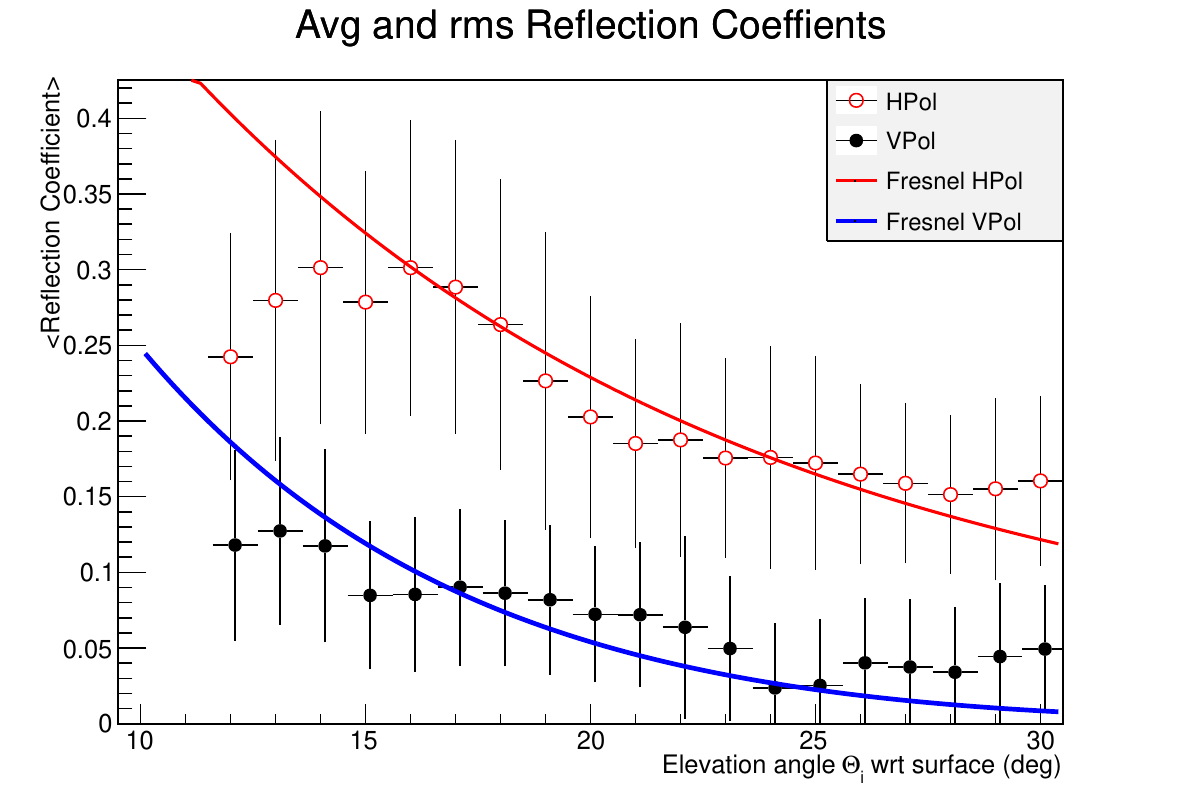}}
\caption{Mean and rms of data presented in previous Figure, as a function of incident elevation angle. For each $\theta_i$ bin, circles represent the average of the points in a given one-degree wide $\theta_i$ slice. Error bars depict the widths (one standard deviation) in each $\theta_i$ slice. The errors on the mean, which are not shown, are obviously considerably smaller than the errors shown.}
\label{fig:FF1.pdf}
\end{figure}
\clearpage

\begin{figure}[htpb]
\centerline{\includegraphics[width=0.6\textwidth]{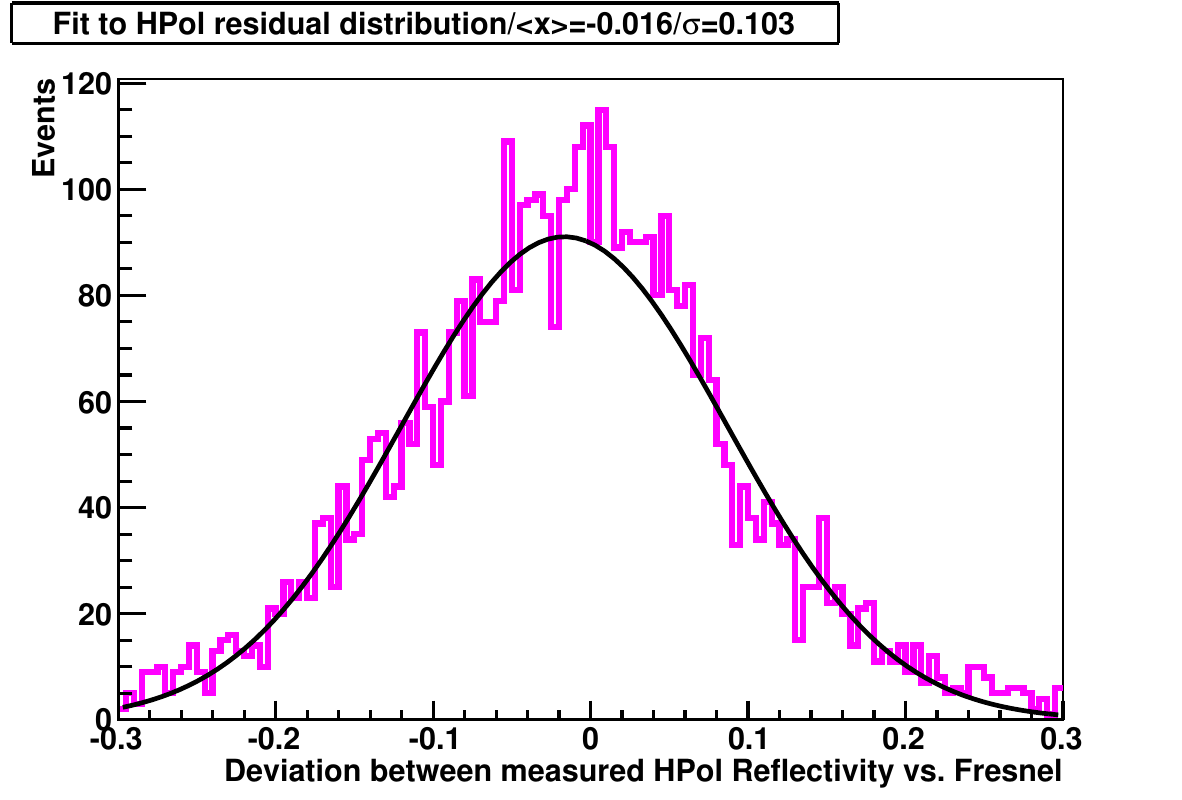}}
\caption{Projection of HPol deviation from Fresnel expectation, with fit to Gaussian distribution overlaid. Fitted mean of distribution is found to be very close to zero (as shown on plot).}
\label{fig:ResidualFit.pdf}
\end{figure}
\clearpage
\begin{figure}[htpb]
\centerline{\includegraphics[width=0.6\textwidth]{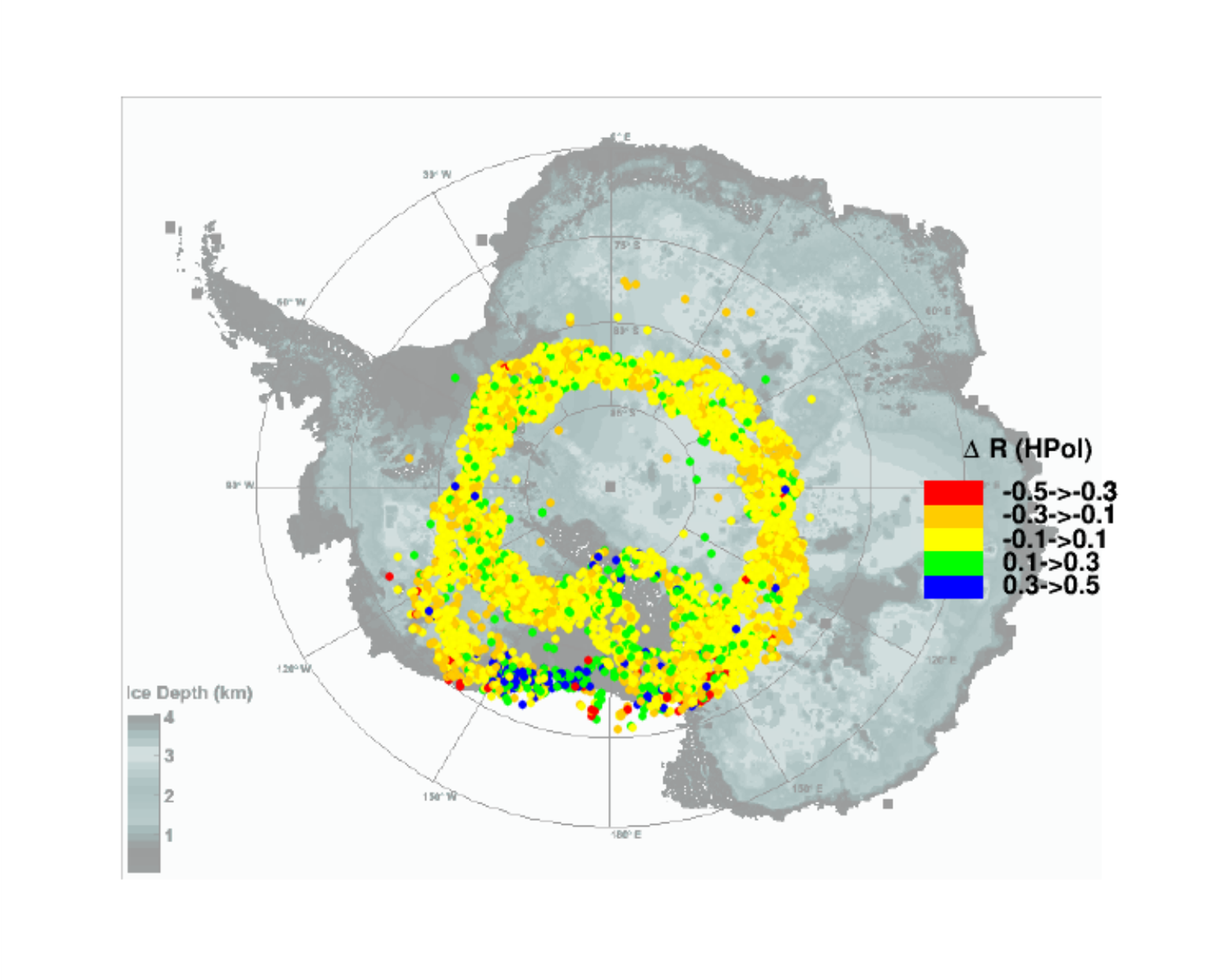}}
\caption{Deviation between measured HPol reflection coefficient relative to reflection coefficient expected at an air-ice sheet interface (from the Fresnel equations), as a function of position on the Antarctic continent. Key in bottom left refers to ice thickness at a given latitude and longitude; ice thickness values are taken from BEDMAP. ANITA's sensitivity is greatest for those portions of the ice sheet which are thickest. We note the largest deviation, relative to the simple-minded expectation for ice, at those times when ANITA is viewing Solar reflections over water, for which the Fresnel coefficients should be markedly larger than for ice.}
\label{fig:HIceWater}
\end{figure}
\begin{figure}[htpb]
\centerline{\includegraphics[width=0.6\textwidth]{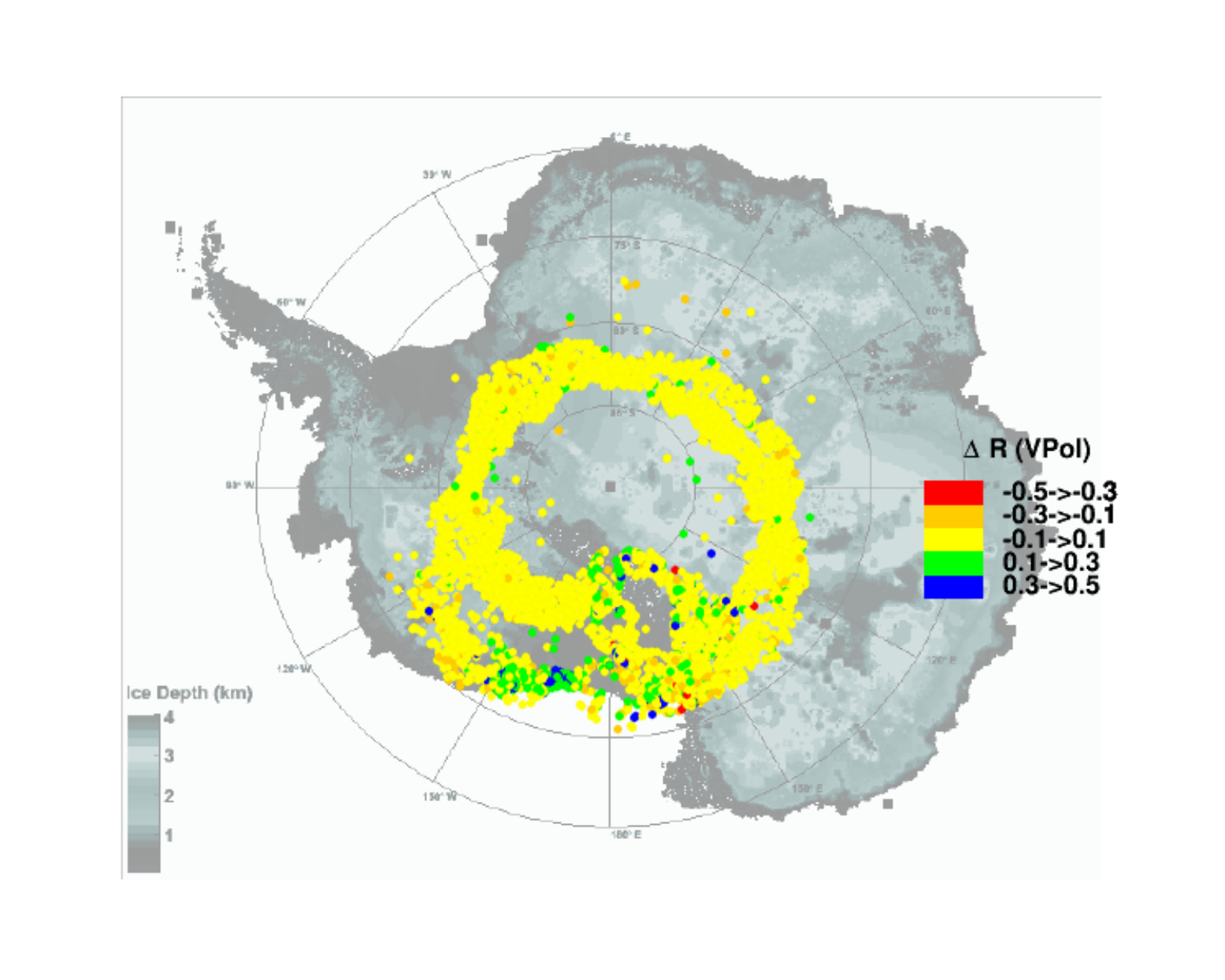}}
\caption{Analogous to previous plot, deviation between measured VPol reflection coefficient relative to reflection coefficient expected at an air-ice sheet interface (from the Fresnel equations), as a function of position on the Antarctic continent. Ice thickness values are taken from BEDMAP.}
\label{fig:VIceWater}
\end{figure}
\begin{figure}[htpb]
\centerline{\includegraphics[width=0.6\textwidth]{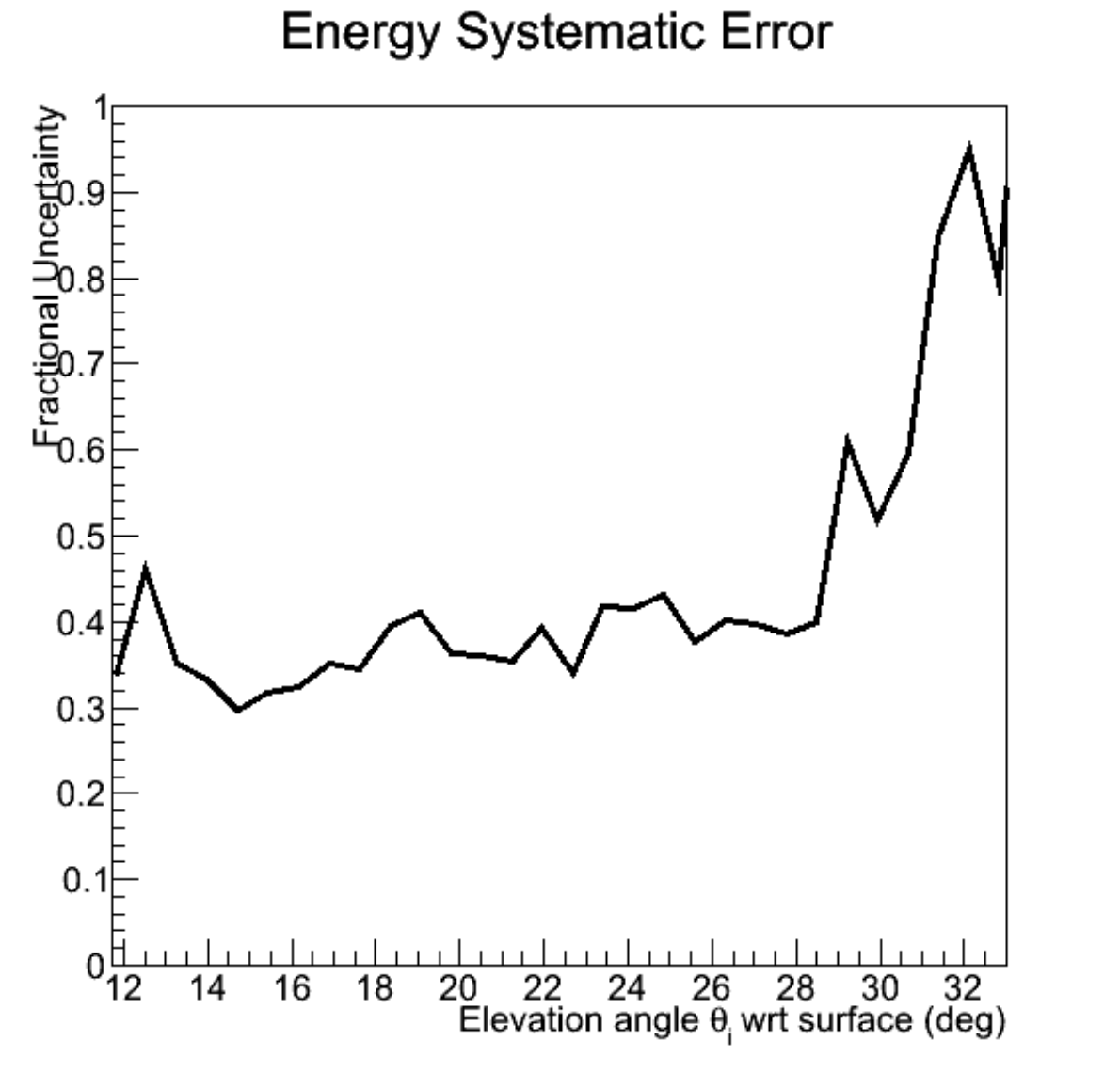}}
\caption{Implied event-by-event radio air shower energy uncertainty, assuming a) errors are equivalent to those obtained in current analysis, and b) energy estimate is dominated by HPol response. Values are equivalent to relative error bars shown in Figure \ref{fig:FF1.pdf}, rather than observed deviation between measurement and Fresnel expectation.}
\label{fig:dErg}
\end{figure}

%
%


\end{document}